\documentclass[a4paper,fleqn,usenatbib]{mnras}

\usepackage[T1]{fontenc}
\usepackage{ae,aecompl}

\usepackage{graphicx}	
\usepackage{amsmath}	
\usepackage{amssymb}	

\newcommand{\kms}{\, {\rm km\,s^{-1}}}
\newcommand{\hMpc}{\,h^{-1}\, {\rm Mpc}}
\newcommand{\hk}{\,h\,{\rm Mpc^{-1}}}

\newcommand{\VEC}{\mathbfit}

\title[kSZ effect in Fourier space]{A direct measure of free electron gas via the kinematic Sunyaev-Zel'dovich effect in Fourier-space analysis}

\author[N. S. Sugiyama et al.]{
Naonori S. Sugiyama$^{1,2}$\thanks{E-mail: nao.s.sugiyama@gmail.com},
Teppei Okumura$^{3,1}$
and David N. Spergel$^{4,5}$
\\
$^{1}$ Kavli Institute for the Physics and Mathematics of the Universe (WPI), \\
       Todai Institutes for Advanced Study, The University of Tokyo, Chiba 277-8582, Japan\\
$^{2}$ CREST, Japan Science and Technology Agency, Kawaguchi, Saitama, Japan \\
$^{3}$ Institute of Astronomy and Astrophysics, Academia Sinica, P. O. Box 23-141, Taipei 10617, Taiwan  \\
$^{4}$ Department of Astrophysical Sciences, Princeton University, Peyton Hall, Princeton NJ 08544-0010, USA \\
$^{5}$ Center for Computational Astrophysics, Flatiron Institute, NY NY 10010, USA
}

\date{}

\pubyear{2017}

\begin{document}
\label{firstpage}
\pagerange{\pageref{firstpage}--\pageref{lastpage}}
\maketitle

\begin{abstract}
	We present the measurement of the kinematic Sunyaev-Zel'dovich (kSZ) effect in Fourier space, rather than in real space. We measure the density-weighted pairwise kSZ power spectrum, the first use of this promising approach, by cross-correlating a cleaned cosmic microwave background (CMB) temperature map, which jointly uses both \textit{Planck} Release 2 and \textit{Wilkinson Microwave Anisotropy Probe} nine-year data, with the two galaxy samples, CMASS and LOWZ, derived from the Baryon Oscillation Spectroscopic Survey (BOSS) Data Release 12. To estimate the CMB temperature distortion associated with each galaxy, we apply an aperture photometry filter. With the current data, we constrain the average optical depth $\tau$ multiplied by the ratio of the Hubble parameter at redshift $z$ and the present day, $E=H/H_0$; we find $\tau E = (3.95\pm1.62)\times10^{-5}$ for LOWZ, which corresponds to the statistical significance of ${\rm S/N}=2.44$, and $\tau E = (1.25\pm 1.06)\times 10^{-5}$ for CMASS, which is consistent with a null hypothesis of no signal. While this analysis results in the kSZ signals with only evidence for a detection,  the combination of future CMB and spectroscopic galaxy surveys should enable precision measurements. We estimate that the combination of CMB-S4 and data from Dark Energy Spectroscopic Instrument should yield detections of the kSZ signal with ${\rm S/N}=70\,\mathchar`-\,100$, depending on the resolution of CMB-S4.
\end{abstract}

\begin{keywords}
cosmology: cosmic background radiation -- cosmology: large-scale structure of Universe -- 
cosmology: observations -- cosmology: theory -- galaxies: intergalactic medium
\end{keywords}

\section{INTRODUCTION}
\label{Sec:Introduction}

\citet{Sunyaev1970kSZ} showed  that a cloud of moving free electrons  induces brightness temperature anisotropies in the cosmic microwave background (CMB) radiation: the amplitude of the \textit{kinematic} Sunyaev-Zel'dovich (hereafter, kSZ) is proportional to the product of the electron column density and the electron velocity. The kSZ signal thus offers a unique and powerful observational tool for a direct measurement of baryonic matter that is insensitive to the free electron gas temperature. The kSZ effect directly probes the difficult to detect ionized warm and hot intracluster medium and the intergalactic medium that is believed to contain most of the universe's baryons (see e.g.~\citet{Fukugita2004}). Because the signal is sensitive to the peculiar velocity probe, it is potentially a powerful probe of large-scale velocity fields and cosmological parameters.

Based on the \textit{pairwise} kSZ estimator proposed by~\citet{Ferreira1999}, \cite{Hand2012} reported the first clear detection of the kSZ effect using CMB data from the Atacama Cosmology Telescope (ACT;~\citet{Swetz2011}) with galaxy positions from the Baryon Oscillation Spectroscopic Survey (BOSS;~\citet{Eisenstein2011}), four decades after the theory of the effect was first proposed. Since this first detection several other detections have followed from multiple CMB experiments, cross-correlating with other galaxy or cluster catalogues: the Planck collaboration~\citep{Planck2016A&A...586A.140P} used galaxy positions from the Sloan Digital Sky Survey Data Release 7 (SDSS-DR7;~\citet{Abazajian2009ApJS..182..543A}), the South Pole Telescope (SPT;~\citet{George2015ApJ...799..177G}) collaboration~\citep{Soergel2016MNRAS.461.3172S} used cluster positions from the Dark Energy Survey (DES;~\citet{DES2005astro.ph.10346T,DES2016MNRAS.460.1270D}), and \citet{DeBernardis2016arXiv160702139D} used CMB data from ACT with galaxy positions from BOSS. In addition to these kSZ measurements, \cite{Schaan2016} reported a detection of the kSZ signal by stacking the CMB temperature at the location of each halo, weighted by the corresponding reconstructed velocity, using CMB data from ACTPol~\citep{Naess2014JCAP...10..007N} with galaxy positions from BOSS. \cite{Hill2016PhRvL.117e1301H} and \cite{Ferraro:2016ymw} detected the kSZ effect through a three point correlation using CMB data from Planck and Wilkinson Microwave Anisotropy Probe (WMAP;~\citet{Bennett2013ApJS..208...20B}) with galaxy positions from the Wide-field Infrared Survey Explorer (WISE;~\citet{Wright2010AJ....140.1868W}).

The kSZ effect is well-suited for studying properties of the optical depth, $\tau$, of haloes hosting galaxies or galaxy clusters. As the measured optical depth via the kSZ effect is insensitive to gas temperature and redshift, the kSZ effect can be used to detect ionized gas that is difficult to observe through its emission, so-called ``missing baryons`` (e.g., \cite{Fukugita1998ApJ...503..518F,Fukugita2004,Bregman2007ARA&A..45..221B}). Most of the missing baryons are thought to reside in a tenuous warm-hot intergalactic medium (WHIM) within a temperature range of $10^5\, {\rm K} < T < 10^7\, {\rm K}$~\citep{Cen1999ApJ...514....1C,Dave1999ApJ...511..521D,Cen2006}, which are neither hot enough (T < $10^{8}\,$K) to be seen in X-ray observations, nor cold enough (T > $10^{3}\,$K) to be made into stars and galaxies. Constraining the optical depth will allow one to investigate astrophysical effects such as star formation and feedback from active galactic nuclei and supernovae for a wide mass and redshift range of the haloes sourcing the kSZ signal~\citep{DeDeo2005,Bregman2007ARA&A..45..221B,Hernandez2008,Ho2009b,Hernandez2009,Shao2011,Hernandez2015,Schaan2016,Flender2016,Battaglia2016JCAP...08..058B,Flender2016arXiv161008029F}.

From a cosmological perspective, the kSZ effect is a potential proxy for the peculiar velocity of galaxies or galaxy clusters. The pairwise kSZ signal enables us to measure the \textit{pairwise velocity}~\citep{Peebles1980lssu.book.....P} at cosmological distances. This large-scale velocity information provides constraints on cosmological parameters, especially on the equation of state of dark energy and modified gravity models~\citep{DeDeo2005,Hernandez2006,Bhattacharya2007ApJ...659L..83B,Bhattacharya2008, Kosowsky2009PhRvD..80f2003K, Keisler2013,Ma2014,Mueller2015a,Alonso2016PhRvD..94d3522A,Sugiyama2016arXiv160606367S} as well as the sum of the neutrino masses~\citep{Mueller2015b}.

In~\citet{Sugiyama2016arXiv160606367S}, we proposed a new statistic of the kSZ signal, the density-weighted pairwise kSZ power spectrum in Fourier space, while many previous works focused on the pair-weighted pairwise kSZ signal in configuration space. There, we showed that two effects, redshift space distortions (RSD)~\citep{Kaiser1987MNRAS.227....1K} and the Alcock-Paczy\'nski test~\citep{Alcock1979Natur.281..358A}, leave their distinctive features on the pairwise kSZ power spectrum, allowing us to infer cosmology from the joint analysis of the density and kSZ power spectra independent of our knowledge of the optical depth in a given tracer sample.
 
This work reports the first measurement of the pairwise kSZ power spectrum using a cleaned CMB map~\citep{Bobin:2015fgb}, which jointly uses both Planck and WMAP data, with two samples of BOSS galaxies, LOWZ and CMASS, drawn from the final Data Release 12 (DR12; \citet{Alam2015ApJS..219...12A}). We improve on previous, similar analyses in a number of ways. First, we use a theoretical model for the pairwise kSZ power spectrum including the RSD effect, which is developed in~\citet{Sugiyama2016arXiv160606367S}. Secondly, we measure the ``density-weighed'' pairwise kSZ estimator normalized by a random catalogue, while almost all previous works measure the ``pair-weighted'' pairwise kSZ estimator normalized by an observed catalogue itself~\citep{Hand2012,Planck2016A&A...586A.140P,Soergel2016MNRAS.461.3172S,DeBernardis2016arXiv160702139D}. This approach allows us to perform the analysis in Fourier space, namely measure the pairwise kSZ power spectrum, similarly to the galaxy power spectrum in clustering analysis. Thirdly, we use a random catalogue to both define number density fluctuations of galaxies and measure survey window functions, correcting for survey geometry effects. Fourth, to break a strong degeneracy among the optical depth $\tau$, the linear logarithmic growth rate $f$, and the linear bias $b$ in the pairwise kSZ signal, we fix the values of $f$ and $b$, where they are measured from the RSD analysis of galaxy clustering~\citep{Gil-Marn2016MNRAS.460.4188G} using the same galaxy samples as we use in this work. In this way, we do not constrain the optical depth $\tau$ itself but the product $\tau E$, where $E=H/H_0$, and $H$ and $H_0$ are the Hubble parameter at given redshift and the present day, respectively. Finally, we repeat the pairwise kSZ power analysis for various angular radii of an aperture photometry filter, where the filter is used to estimate the CMB temperature distortion associated with each galaxy. We measure the optical depth $\tau E$ as a function of aperture radii through this analysis and constrain parameters in an assumed gas profile as well as the gas-mass fraction via the measurement.

We provide a forecast of future kSZ measurements and their power to constrain optical depth models with an error model for the optical depth. We assume a hypothetical Stage-IV CMB experiment (CMB-S4;~\cite{Abazajian2015APh....63...55A}) with one current galaxy survey, BOSS, and two upcoming spectroscopic galaxy surveys, the Dark Energy Spectroscopic Instrument (DESI;~\cite{Levi2013arXiv1308.0847L}) and the Subaru Prime Focus Spectrograph (PFS;~\cite{Takada2014PASJ...66R...1T}).

This paper is organized as follows. In Section~\ref{Sec:Data}, we summarize the CMB and galaxy sample data used in our analysis. In Section~\ref{Sec:Theory}, we describe our models for both the optical depth and the pairwise kSZ power spectrum. In Section~\ref{Sec:Methodology}, we explain the technique to measure the pairwise kSZ power spectrum, followed by the treatment of the survey window function in Section~\ref{Sec:SurveyWindowFunction}. Section~\ref{Sec:Covariance} presents the method to estimate the covariance matrix of the pairwise kSZ power spectrum from data itself. In Section~\ref{Sec:Results}, the results including the best-fitting parameters of the filtered optical depth are presented. We give an overview of future kSZ measurements in Section~\ref{Sec:Forecasts}. We present a summary and conclusions in Section~\ref{Sec:Conclusions}. We additionally provide Appendix~\ref{Ap:Covariance} that compute the pairwise kSZ power covariance in linear theory and compare it with the covariance estimates from data itself,
Appendix~\ref{Ap:Details} that details how to measure the pairwise kSZ power spectrum 
and Appendix~\ref{Ap:Derivations} that gives detailed derivations of equations used in our analysis.

Throughout this paper we adopt a flat $\Lambda$CDM cosmology~\citep{Planck2016A&A...594A..13P}: $\Omega_{\rm m}=0.309$, $\Omega_{\rm \Lambda} = 0.691$, $n_{\rm s} = 0.9608$, $\sigma_8=0.815$, and $H_0 = 100\, h\, {\rm km\, s^{-1}\, Mpc^{-1}}$ with $h=0.68$. We use the best-fitting values of $f\sigma_8$ and $b\sigma_8$ measured in~\citep{Gil-Marn2016MNRAS.460.4188G}: $(f\sigma_8,\, b\sigma_8)=(0.392,\,1.283)$ for LOWZ and $(0.445,\,1.218)$ for CMASS.

\section{DATA}
\label{Sec:Data}

\subsection{ \textit{WMAP} and \textit{Planck} maps}
\label{Sec:Data_CMB}

We use a cleaned CMB temperature map constructed from a novel component separation technique, ``local-generalized morphological component analysis'' (LGMCA;~\citet{Bobin2013a&a...550a..73b,Bobin2014A&A...563A.105B}) that uses both \textit{Planck} Release 2 (PR2; ~\citet{Planck2016A&A...594A...1P}) and \textit{Wilkinson Microwave Anisotropy Probe} nine-year (\textit{WMAP9}; \citet{Bennett2013ApJS..208...20B}) data, where the PR2 and \textit{WMAP9} maps contain nine and five channels, respectively. This WPR2 map is provided in \textit{HEALPix}~\citep{Gorski2005ApJ...622..759G} pixelization scheme at a resolution of the grid expressed by $N_{\rm side}=2048$.

The WPR2 map is well suited for an analysis of the kSZ signal, as it is designed to have minimal galactic contamination and very little residual thermal Sunyaev-Zel'dovich (tSZ) signal. In this map, residual contamination of the tSZ is about $10^6$ times smaller than the kSZ signal at any scale (see fig. $12$ in \cite{Bobin2014A&A...563A.105B}). We also use the WPR2 noise map made by applying the LGMCA technique to the Planck and WMAP noise maps, in order to perform a null test of the kSZ signal as a test of systematics (see Section~\ref{Sec:Results}).

\subsection{BOSS galaxy catalogues}
\label{Sec:Data_Galaxy}

As proxies for haloes, we use two galaxy samples, the  LOWZ sample with $463044$ galaxies between $z = 0.15\mathchar`-0.43$ $(z_{\rm eff} = 0.33)$, and the CMASS sample with $849637$ galaxies between $z = 0.43\mathchar`-0.7$ $(z_{\rm eff} = 0.56)$.  These samples are   drawn from  the Baryon Oscillation Spectroscopic Survey (BOSS;~\citet{Eisenstein2011, Bolton2012AJ....144..144B,Dawson2013AJ....145...10D}) Data Release 12 (DR12; \citet{Alam2015ApJS..219...12A}) and  are selected from multi-colour SDSS imaging~\citep{Fukugita1996,Gunn1998,Smith2002AJ....123.2121S,Gunn2006,Doi2010AJ....139.1628D}.
We also use the associated random catalogues that quantify the survey geometry of BOSS.

The LOWZ galaxies lie in massive haloes with mean halo masses of $5.2\times10^{13}\, h^{-1}\, M_{\sun}$, large-scale bias of $\sim 2.0$ and the satellite fraction of $12\%$, while the CMASS galaxies lie in halos with the masses of $2.6\times10^{13}\, h^{-1}\, M_{\sun}$, the bias $\sim 2.0$ and the satellite fractions $10\%$~\citep{White2011ApJ...728..126W,Parejko2013MNRAS.429...98P,Bundy2015ApJS..221...15B,Leauthaud2016MNRAS.457.4021L,Saito2016MNRAS.460.1457S}. In our analysis, we use the CMASS and LOWZ galaxy samples in both the North Galactic Cap (NGC) and the South Galactic Cap (SGC).

To correct for several observational artefacts in the catalogues and obtain unbiased estimates of galaxy density fields, we use the following weight for each galaxy $i$~\citep{Ross2012MNRAS.424..564R,Anderson2014MNRAS.441...24A,Reid2016MNRAS.455.1553R},
\begin{eqnarray}
	w_{ {\rm c}, i} = w_{ {\rm systot}, i} \left( w_{ {\rm cp}, i} + w_{ {\rm noz}, i} - 1\right),
	\label{Weight}
\end{eqnarray}
where $w_{\rm cp}$, $w_{\rm noz}$, and $w_{\rm systot}$ denote a redshift failure weight, a collision weight, and an angular systematics weight, respectively. The details about the observational systematic weights are described in~\citet{Reid2016MNRAS.455.1553R}. We do not apply the optimal weighting of galaxies, the so-called FKP weight~\citep{Feldman1994ApJ...426...23F}, because this weight is derived for the galaxy power spectrum and would be non-optimal for the pairwise kSZ spectrum.

\section{THEORY}
\label{Sec:Theory}

\subsection{Kinematic Sunyaev-Zel'dovich effect}
\label{Sec:kSZ}

The scattering of CMB photons with electrons moving with respect to the rest frame of CMB, where the CMB is observed to be isotropic, causes anisotropies of the CMB brightness temperature. This bulk motion-induced thermal distortion is known as the kSZ effect. In the non-relativistic limit, the temperature anisotropies due to the kSZ effect is given by~\citep{Sunyaev1970kSZ,Sunyaev1972,Sunyaev1980,Ostriker1986}
\begin{eqnarray}
	\delta T_{\rm kSZ}(\hat{n}) = - T_0\sigma_{\rm T}\int dl \left( \frac{\vec{v}_{\rm e}(\vec{x})\cdot{\hat n}}{c} \right) n_{\rm e}(\vec{x}),
	\label{Eq:kSZ}
\end{eqnarray}
where $T_0=2.725\,K$ is the average CMB temperature, $n_e$ is the physical free electron number density, $\vec{v}_e$ denotes the peculiar velocity of free electrons, $\sigma_{\rm T}$ is the Thomson scattering cross-section, and $c$ is the speed of light. In this expression, the integral $\int dl$ is performed along the line-of-sight (LOS) given by $\hat{n}$. The sign of velocities is defined so that $\vec{v}\cdot\hat{n}>0$ corresponds to a cloud of free electrons moving away from the observer. In this definition, the kSZ effect slightly shifts CMB temperature, lower (higher) for positive (negative) peculiar velocities, with preserving the CMB blackbody spectrum and without depending on free electron temperature.

\subsubsection{Galaxies as point sources}

We start with the simplifying assumption that observed galaxies sit  the centres of their host haloes and can be treated as point sources. Under this assumption, we can describe the electron number density as
\begin{eqnarray}
	n_{\rm e}(\vec{x}) = \frac{1}{a^3}\sum_i^{N} N_{ {\rm e},i} \delta_{\rm D}\left( \vec{x} - \vec{x}_i \right),
\end{eqnarray}
where $N_{ {\rm e }, i}$ means the total number of free electrons associated with galaxy $i$, $N$ is the total number of observed galaxies, $a$ is the scale factor, $\vec{x}_i$ represents the three-dimensional coordinates of the galaxy, and $\delta_{\rm D}$ is the Delta function. Then, equation.~(\ref{Eq:kSZ}) becomes
\begin{eqnarray}
	\delta T_{\rm kSZ}(\hat{n}) = -\sum_i^{N} 
	\left( \frac{\vec{v}_{ {\rm e},i} \cdot \hat{n}_i}{c} \right)
	\frac{T_0\sigma_{\rm T}N_{ {\rm e},i}}{D_{ {\rm A},i}^2} \delta_{\rm D}\left( \hat{n} - \hat{n}_i \right),
	\label{Eq:kSZ_1}
\end{eqnarray}
where we used $\delta_{\rm D}\left( \vec{x}-\vec{x}_i \right) = (1/\chi^2)\delta_{\rm D}\left( \chi - \chi_i \right)\delta_{\rm D}\left( \hat{n}-\hat{n}_i \right)$ and $\int dl = a\int d\chi$ with $\chi$ being the radial comoving distance, the free electron peculiar velocity $\vec{v}_{ {\rm e},i}$ is evaluated at galaxy $i$, the line-of-sight (LOS) direction $\hat{n}_i$ points to the galaxy from the observer, and $D_{ {\rm A},i}=a\chi_i$ represents the angular diameter distance to the galaxy. In subsection \ref{Sec:profiles}, we will discuss the more realistic case where galaxies are not point sources but have gas profiles. 

The observed kSZ temperature $\delta T_{\rm kSZ}^{(\rm obs)}$ can be expressed as the convolution of the true kSZ temperature $\delta T_{\rm kSZ}$ with the instrumental beam function $B(\hat{n},\hat{n}')$,
\begin{eqnarray}
	\delta T_{\rm kSZ}^{(\rm obs)}(\hat{n}) = \int d\Omega_{\hat{n}'}\, B(\hat{n},\hat{n}')\, \delta T_{\rm kSZ}(\hat{n}').
	\label{Eq:kSZ_Beam}
\end{eqnarray}
We approximate the Planck beam as a Gaussian,
\begin{eqnarray}
	\scalebox{0.95}{$\displaystyle
	\delta T_{\rm kSZ}^{(\rm obs)}(\hat{n}) =   -\sum_i^{N} 
	\left( \frac{\vec{v}_{ {\rm e},i} \cdot \hat{n}_i}{c} \right)
	\frac{T_0\sigma_{\rm T}N_{ {\rm e},i}}{2\pi D_{ {\rm A},i}^2 \sigma_{\rm B}^2} \,
	e^{-\left( \hat{n} - \hat{n}_i \right)^2/(2\sigma_{\rm B}^2)},$}
	\label{Eq:kSZ_GaussianBeam}
\end{eqnarray}
where $\sigma_{\rm B}$ is related to the effective beam full-width at half-maximum (FWHM) as $\sigma_{\rm B} = {\rm FWHM}/(\sqrt{8 \ln(2)}) = 0.4247\, {\rm FWHM}$. In our analysis, we use ${\rm FWHM}=5\arcmin$ for the WPR2 map (Section.~\ref{Sec:Data_CMB}).

We apply an aperture photometry (AP) filter to estimate the CMB temperature distortion associated with each galaxy
\begin{eqnarray}
	\delta T_{\rm kSZ}^{(\rm AP)}(\vec{\theta}\,) = \int d^2\theta'\, U(\vec{\theta}-\vec{\theta}\,')\,
	\delta T_{\rm kSZ}^{(\rm obs)}(\vec{\theta}\,'),
	\label{Eq:TAP1}
\end{eqnarray}
where we describe the three-dimensional LOS unit vector $\hat{n}$ as the two-dimensional vector $\vec{\theta}$ in the $\theta_x-\theta_y$ plane and centre each galaxy at the origin. The weight function $U(\vec{\theta}\,)$ satisfies the criterion 
\begin{eqnarray}
	\int_0^{\theta_{\rm c}} d\theta\, \theta\, U(\theta) = 0,
\end{eqnarray}
where $\theta = |\vec{\theta}|$, and $\theta_{\rm c}$ is the AP filter radius. In other words, $U(\theta)$ is taken to be a {\it compensated} radial weight function across the aperture. This filter is insensitive to CMB fluctuations on scales larger than the aperture radius, so that the contamination of primary CMB anisotropies to the kSZ signal in the filtered map is nearly uncorrelated on large scales (see Appendix~\ref{Ap:Covariance}). Using the convolution theorem, equation~(\ref{Eq:TAP1}) becomes
\begin{eqnarray}
	\delta T_{\rm kSZ}^{(\rm AP)}(\vec{\theta}\,) = \int \frac{d^2\ell}{(2\pi)^2}\, e^{i\vec{\ell}\cdot\vec{\theta}}\,
	U(\ell \theta_{\rm c})\, \delta T_{\rm kSZ}(\vec{\ell}\,)\, B(\vec{\ell}\,), 
	\label{AP1}
\end{eqnarray}
where $\vec{\ell}$ is the two-dimensional wavevector perpendicular to the LOS, $\ell = |\vec{\ell}\,|$, and $\delta T_{\rm kSZ}(\vec{\ell}\,)$ and $B(\vec{\ell}\,)$ are the two-dimensional Fourier transforms of the kSZ temperature and the beam function, respectively. From equation~(\ref{Eq:kSZ_1}), $\delta T_{\rm kSZ}(\vec{\ell}\,)$ is given by
\begin{eqnarray}
	\delta T_{\rm kSZ}(\vec{\ell}\,) = -\sum_i^{N} 
	\left( \frac{\vec{v}_{ {\rm e},i} \cdot \hat{n}_i}{c} \right)
	\frac{T_0\sigma_{\rm T}N_{ {\rm e},i}}{D_{ {\rm A},i}^2} e^{-i\vec{\ell}\cdot\vec{\theta}_i}.
	\label{AP2}
\end{eqnarray}
For a compensated top-hat window, we average the mean temperature within a disk of radius $\theta_{\rm c}$, and then subtract from it the mean temperature in a surrounding ring of inner and outer radii, where the outer radius is chosen to be $\sqrt{2} \theta_{\rm c}$ so that the disk and ring have the same area~\citep{Planck2016A&A...586A.140P,Schaan2016,DeBernardis2016arXiv160702139D}. Then, the Fourier transform of the AP filter is given by~\citep{Alonso2016PhRvD..94d3522A}
\begin{eqnarray}
	U(x) = 2\left[W_{\rm top}(x) - W_{\rm top}(\sqrt{2}x) \right]
	\label{Eq:APfilter}
\end{eqnarray}
using the top-hat smoothing window function
\begin{eqnarray}
	W_{\rm top}(x) = 2 \frac{J_1(x)}{x},
	\label{TopHat}
\end{eqnarray}
where $J_1$ is the first Bessel function of the first kind.

We assume that the AP filtered kSZ temperature $\delta T_{\rm kSZ}^{\rm (AP)}$ associated with galaxy $i$ is not affected by the other galaxies but only by the galaxy itself. Then, using equations~(\ref{AP1}) and (\ref{AP2}), we obtain
\begin{eqnarray}
	\delta T_{\rm kSZ}^{(\rm AP)}(\hat{n}_i) \simeq - \frac{T_0 \tau_i}{c}\, \vec{v}_{ {\rm e}, i}\cdot\hat{n}_i,
\end{eqnarray}
where the optical depth $\tau_i$ estimated at galaxy $i$ is given by
\begin{eqnarray}
	\tau_i = \frac{\sigma_{\rm T}N_{ {\rm e},i}}{D_{ {\rm A},i }^2}
	\int \frac{d^2\ell}{(2\pi)^2}\, U(\ell\theta_{\rm c})\, B(\vec{\ell}\,).
	\label{Eq:tau_i_1}
\end{eqnarray}

We simply take the gas-traces-mass assumption and admit the universalities of the gas-mass fraction $f_{\rm gas}$ and the host halo mass $M$ for galaxies, resulting in $N_{{\rm e}}\sim f_{\rm gas}M/(\mu_{\rm e}m_{\rm p})$, where $\mu_{\rm e}=1.17$ is the mean particle weight per electron, and $m_{\rm p}$ is the proton mass. We estimate the angular diameter distance at effective redshift $z_{\rm eff}$ of a given galaxy sample. Then, the optical depth becomes the approximately-same at each galaxy: $\tau = \tau_i$, where $\tau$ is interpreted as an ``average'' optical depth that is a proportionality constant to fit the observed kSZ temperature to the prediction of the LOS peculiar velocity. This estimation corresponds to the assumption that there is no strong correlation between the optical depth and the peculiar velocity for a given galaxy (halo). In addition, we assume that the peculiar velocity of free electrons can be regarded as the velocity of dark matter: $\vec{v} = \vec{v}_{\rm e}$. Under the above assumptions, the relation between the kSZ temperature and the LOS peculiar velocity simplifies to
\begin{eqnarray}
	  \delta T^{(\rm AP)}_{\rm kSZ}(\hat{n}_i) = -\frac{T_0\tau}{c}\, \vec{v}_i\cdot\hat{n}_i,
     \label{Eq:kSZ_tau}
\end{eqnarray}
where the average optical depth is given by
\begin{eqnarray}
	  \tau =  \frac{\sigma_{\rm T} f_{\rm gas} M}{\mu_{\rm e} m_{\rm p}D_{{\rm A}}^2(z_{\rm eff})}
	  \int \frac{d^2\ell}{(2\pi)^2}\, U(\ell\theta_{\rm c})\, B(\vec{\ell}\,).
	\label{tau0}
\end{eqnarray}
Most of the electrons in the halo are in ionized gas and   are traced by the kSZ effect.  We expect that only a small fraction of the mass is in neutral gas and stars\citep{Fukugita2004}. Therefore, we expect that the value of the gas fraction $f_{\rm gas}$ is similar to the universal baryon fraction $f_{\rm b}=\Omega_{\rm b}/\Omega_{\rm m}=0.155$, which is tightly constrained at high redshift by measurements of CMB~\citep{Hinshaw2013ApJS..208...19H,Planck2016A&A...594A..13P} and of the abundance of light elements formed through the process of Big Bang nucleosynthesis (BBN)~\citep{Steigman2007ARNPS..57..463S}.

\subsubsection{Realistic Gas profiles}
\label{Sec:profiles}

We now relax the point source assumption and allow the gas in galaxies to have an extended profile. However, for the LOWZ and CMASS halos the exact shape of the gas profile that induces the kSZ effect is currently very uncertain. In this work, we assume that on average the gas profile can be expressed as a projected gas profile $N(\vec{\theta}\,)$, and that the kSZ temperature is expressed as the convolution of the point source kSZ temperature with an assumed gas profile function. Then, we finally derive the optical depth model assuming the gas-profile
\begin{eqnarray}
	\tau =  \frac{\sigma_{\rm T} f_{\rm gas} M}{\mu_{\rm e} m_{\rm p}D_{{\rm A}}^2(z_{\rm eff})}
	\int \frac{d^2\ell}{(2\pi)^2}\, U(\ell\theta_{\rm c})\, N(\vec{\ell}\, )\, B(\vec{\ell}\,),
	\label{tau1}
\end{eqnarray}
where $N(\vec{\ell}\,)$ is the Fourier transform of an assumed projected gas profile.
In the following analysis we consider two functional forms for the projected gas profiles, (1) a Gaussian profile with a characteristic radius, $\sigma_{\rm R}$~\citep{Schaan2016} and (2) a $\beta$-profile with slope $\beta$ and core radius $\theta_{\rm R}$~\citep{Cavaliere1976A&A....49..137C}, given by
\begin{eqnarray}
	N_{\rm G}(\theta) &=& \frac{1}{2\pi \sigma_{\rm R}^2} e^{-\theta^2/(2\sigma_{\rm R}^2)} \nonumber \\
	N_{\beta}(\theta) &=& N_0\left( 1 + \frac{\theta^2}{\theta_{\rm R}^2} \right)^{-\frac{3}{2}\beta+\frac{1}{2}},
	\label{Eq:gas_profile}
\end{eqnarray}
where $N_0$ is the normalization factor so that $\int d^2\theta N_{\beta}(\theta) = 1$. Furthermore, we use the Gaussian beam function, $B(\ell) = e^{-\sigma_{\rm B}^2\ell^2/2}$.

For the Gaussian gas profile, we can derive a simple analytic form of the optical depth as follows
\begin{eqnarray}
	  \tau =  \frac{\sigma_{\rm T} f_{\rm gas} M}{2\pi\mu_{\rm e} m_{\rm p} \Sigma^2 D_{{\rm A}}^2(z_{\rm eff})}
\end{eqnarray}
with
\begin{eqnarray}
	\Sigma^2 = 
	\frac{(4\sigma_{\rm B}^2+4\sigma^2_{\rm R}+\theta_{\rm c}^2)(2\sigma_{\rm B}^2+2\sigma^2_{\rm R} + \theta_{\rm c}^2)}{4\theta_{\rm c}^2},
\end{eqnarray}
where we assumed that the top-hat smoothing window function $W_{\rm top}(x)$ (equation~\ref{TopHat}) is a Gaussian function $W_{\rm top}(x)\sim e^{-(x/2)^2/2}$. We note here that the optical depth has the maximum value at an aperture radius of
\begin{eqnarray}
	\theta_{\rm peak} = 2^{3/4} \sqrt{\sigma_{\rm B}^2 + \sigma_{\rm R}^2}.
	\label{Eq:Peak_position}
\end{eqnarray}
At the peak, we obtain
\begin{eqnarray}
	\Sigma^2 = 2.914\, (\sigma_{\rm B}^2 + \sigma_{\rm R}^2),
\end{eqnarray}
and we find that the amplitude of the kSZ signal is reduced by a factor of $3$ due to the AP filtering, because we obtain $\Sigma^2=\sigma_{\rm B}^2+\sigma_{\rm R}^2$ if we do not apply the filtering. For an optimal aperture radius computed by equation~(\ref{Eq:Peak_position}), we can estimate a typical value of the optical depth as follows
\begin{eqnarray}
	\tau &=&  5.37\times10^{-5}\left( \frac{f_{\rm gas}}{0.155} \right) \left( \frac{M}{10^{14}\,h^{-1}\, M_{\sun}} \right) \nonumber \\
	 &\times& \hspace{-0.2cm}\left( \frac{h}{0.68} \right) 
	\left( \frac{3\arcmin}{\sqrt{\sigma_{\rm B}^2 + \sigma_{\rm R}^2}} \right)^2 \left( \frac{10^3\hMpc}{D_{\rm A}} \right)^2.
	\label{Eq:RoughTau}
\end{eqnarray}
A typical value of the characteristic radius $\sigma_{\rm R}$ in the Gaussian gas profile can be estimated from the halo radius $R$ divided by the angular diameter distance $D_{\rm A}$: $\sigma_{\rm R} = R / D_{\rm A}$. A way of estimating the halo radius $R$ through the halo mass $M$ is 
\begin{eqnarray}
	  R_{\Delta} = \left( \frac{3}{4\pi} \frac{M}{\Delta\, \rho_{\rm crit}(z)} \right)^{1/3},
	  \label{Eq:R_Delta}
\end{eqnarray}
where $R_{\Delta}$ is the radius within which the average density of the halo is $\Delta$ times larger than the critical density $\rho_{\rm crit}(z)$ at redshift $z$. When we choose $\Delta = 200$, we estimate the mean halo radii of $R_{200} = 0.54\hMpc$ for LOWZ and $R_{200} = 0.39\hMpc$ for CMASS, which respectively correspond to $\sigma_{\rm R} = 2.7\arcmin$ and $1.4\arcmin$. Since the beam size for the WPR2 map is $\sigma_{\rm B}=2.1\arcmin$, the LOWZ halos can be resolved, while the CMASS halos cannot. The peak position of the AP filtered optical depth is estimated as $\theta_{\rm peak} = 4.9\arcmin$ for LOWZ and $\theta_{\rm peak} = 4.0\arcmin$ for CMASS. It is worth noting that the Gaussian profile includes the point source assumption in the limit of $\sigma_{\rm R}\to 0$, leading to a peak position of $\theta_{\rm peak}=3.6\arcmin$ for both LOWZ and CMASS. From equations~(\ref{Eq:RoughTau}) and (\ref{Eq:R_Delta}), we find that for $\sigma_{\rm B} \gg \sigma_{\rm R}$ the optical depth is proportional to the mean halo mass $M$ divided by the square of the angular diameter distance $(D_{\rm A}(z))^2$, while for $\sigma_{\rm B} \ll \sigma_{\rm R}$ the optical depth is proportional to $M^{1/3}$ multiplied by $(\rho_{\rm crit}(z))^{2/3}$. 
In other words, the optical depth decreases with increasing redshift for the the former case ($\sigma_{\rm B} \gg \sigma_{\rm R}$), 
while it increases for the latter case ($\sigma_{\rm B}\ll\sigma_{\rm R}$).
Thus, we expect that future kSZ surveys with high-resolution CMB experiments that satisfy $\sigma_{\rm B}\ll\sigma_{\rm R}$
such as Advanced ACTPol~\citep{Calabrese:2014gwa,Henderson:2015nzj} and CMB-S4~\citep{Abazajian2015APh....63...55A}
will allow us to detect high-redshift kSZ signals. (See Section~\ref{Sec:Forecasts} for details.)

\begin{figure}
	\includegraphics[width=\columnwidth]{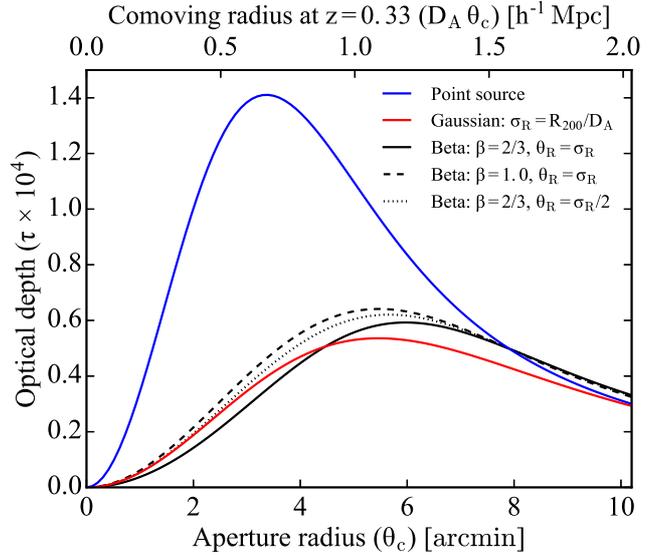}
	\caption{
Optical depth as a function of the angular radius $\theta_{\rm c}$ of the AP filter for various gas profiles (equation~\ref{tau1}). We use the parameters that correspond to the LOWZ sample and the WPR2 map (Sec.~\ref{Sec:Data}): $M=5.2\times 10^{13}\, h^{-1}\,M_{\sun}$, $z=0.33$, ${\rm FWHM}=5\arcmin$, $h=0.68$, $\sigma_{\rm R}=R_{200}/D_{\rm A}$, and $f_{\rm gas}=0.155$. The optical depth amplitudes computed by both the Gaussian profile (red solid) and the $\beta$-profile with different three parameters (black solid, black dashed, black dotted) are smaller than that computed by the point source assumption, and the peak positions predicted by the gas profiles shift outward compared to the point source prediction.
	}
	\label{fig:tau_filter_gaussian_beta}
\end{figure}

For the $\beta$-profile, the integral in the normalization factor $1/N_0=\int d^2\theta \left( 1 + \theta^2/\theta_{\rm R}^2 \right)^{-\frac{3}{2}\beta+\frac{1}{2}}$ has a logarithmic divergence and needs to be truncated; therefore, the optical depth predicted from the $\beta$-profile significantly depends on the upper limit of the integral $\theta_{\rm max}$. For a fair comparison with the Gaussian profile, we choose $\theta_{\rm max}=2 \sigma_{\rm R}$, because $95$ percent of the total number of electrons associated with a given halo is distributed within $2\sigma_{\rm R}$ in the Gaussian profile.

Fig.~\ref{fig:tau_filter_gaussian_beta} shows the optical depth as a function of aperture radius $\theta_{\rm c}$ for various gas profile models~\citep[see also][Fig. 1]{Schaan2016}. We use equation~(\ref{tau1}) with the following parameters that correspond to the LOWZ sample and the WPR2 map: $M=5.2\times 10^{13}\, h^{-1}\,M_{\sun}$, $z=0.33$, ${\rm FWHM}=5\arcmin$, $h=0.68$, $\sigma_{\rm R}=R_{200}/D_{\rm A}$, and $f_{\rm gas}=0.155$. For small $\theta_{\rm c}$, the kSZ signal contributes to the disk and the ring of the AP filter, which leads to a cancellation. For large $\theta_{\rm c}$, the kSZ signal is entirely included in the disk of the AP filter, and the AP filtered optical depth has a peak, depending on the CMB resolution $\sigma_{\rm B}$ and the gas radius $\sigma_{\rm R}$ in equation~(\ref{Eq:Peak_position}). The amplitudes of the optical depth predicted from the Gaussian- and $\beta$-profiles (red and black lines) are smaller than that computed from the point source assumption (blue line), and the peak positions predicted by the gas profiles shift outward compared to the point source prediction.

\subsection{Pairwise kSZ power spectra and correlation functions}
\label{Sec:kSZ_Power}

In this work, we apply the kSZ power spectrum approach developed in \citep{Sugiyama2016arXiv160606367S}. We compute the kSZ power spectrum on the aperture-averaged data,
\begin{eqnarray}
	\scalebox{0.90}{$\displaystyle
	P_{\rm kSZ}(\vec{k}) =  
	\left\langle - \frac{V}{N^2} \sum_{i,j} \left[ \delta T^{(\rm AP)}_{\rm kSZ}(\hat{n}_i) - \delta T^{(\rm AP)}_{\rm kSZ}(\hat{n}_j)\right] 
	e^{-i\vec{k}\cdot\vec{s}_{ij}} \right\rangle $},
	  \label{Eq:PairwiseKSZPower}
\end{eqnarray}
where $\vec{s}_i$ denotes the position of galaxy $i$ in the survey, $\vec{s}_{ij} = \vec{s}_i - \vec{s}_j$, $N$ represents the total number of observed galaxies, and $V$ is the survey volume. The redshift-space position, $\vec{s}_i$, is displaced from the real-space galaxy coordinate, $x_i$, by the velocity along the LOS:
\begin{eqnarray}
	\vec{s}_i = \vec{x}_i + \frac{\vec{v}_i\cdot\hat{n}_i}{aH}\hat{n}_i,
	\label{x_to_s}
\end{eqnarray}
where $\hat{n}_i$ is the LOS, and $H$ is the Hubble parameter. The pairwise kSZ power spectrum is the Fourier transform of the pairwise kSZ correlation function:
\begin{eqnarray}
	\scalebox{0.85}{$\displaystyle
	\xi_{\rm kSZ}(\vec{s}\,) 
	= \left\langle -\frac{V}{N^2} \sum_{i,j} \left[ \delta T^{(\rm AP)}_{\rm kSZ}(\hat{n}_i) - \delta T^{(\rm AP)}_{\rm kSZ}(\hat{n}_j)\right] 
	\delta_{\rm D}\left( \vec{s} - \vec{s}_{ij} \right) \right\rangle $}.
\end{eqnarray}

One can see from equation~(\ref{Eq:kSZ_tau}) that $P_{\rm kSZ}$ and $\xi_{\rm kSZ}$ are respectively proportional to the LOS pairwise velocity power spectrum, $P_{\rm pv}$, and correlation function, $\xi_{\rm pv}$, 
\begin{eqnarray}
	  P_{\rm kSZ}(\vec{k}) &\simeq& \left( \frac{T_0 \tau}{c} \right) P_{\rm pv}(\vec{k}), \nonumber \\
	\xi_{\rm kSZ}(\vec{s}\,)&\simeq& \left( \frac{T_0 \tau}{c} \right) \xi_{\rm pv}(\vec{s}\,),
	\label{kSZ_pk}
\end{eqnarray}
where 
\begin{eqnarray}
	  P_{\rm pv}(\vec{k}) \hspace{-0.25cm}
	  &=& \hspace{-0.25cm}
	  \left\langle \frac{V}{N^2} \sum_{i,j} \left[ \vec{v}_i \cdot \hat{n}_i - \vec{v}_j \cdot \hat{n}_j\right]
	  e^{-i\vec{k}\cdot\vec{s}_{ij}} \right\rangle ,\nonumber \\
	  \xi_{\rm pv}(\vec{s}\,) \hspace{-0.25cm}
	  &=&\hspace{-0.25cm}
	  \left\langle \frac{V}{N^2} \sum_{i,j} \left[ \vec{v}_i \cdot \hat{n}_i - \vec{v}_j \cdot \hat{n}_j\right]
	  \delta_{\rm D}\left( \vec{s} - \vec{s}_{ij} \right)\right\rangle. 
\end{eqnarray}
The LOS pairwise velocity power spectrum and correlation function can be also defined using the density-weighted velocity field of galaxies, which is often referred to as the ``momentum field'' (e.g., \cite{Park2000MNRAS.319..573P}). The LOS momentum field is defined as $p(\vec{x}\,) = [\vec{v}(\vec{x}\,)\cdot\hat{n}](1 + \delta(\vec{x}\,))$ in real space, where $\delta$ is the density contrast, and the translation of $p(\vec{x})$ from real space to redshift space is given by \citep{Okumura2014JCAP...05..003O,Sugiyama:2015dsa}
\begin{eqnarray}
	\scalebox{0.90}{$\displaystyle
	p_{\rm s}(\vec{s}\,) = \int d^3x\, [\vec{v}(\vec{x}\,)\cdot\hat{n}] \left( 1 + \delta(\vec{x}\,) \right)
	\delta_{\rm D}\left( \vec{s} - \vec{x} - \frac{\vec{v}(\vec{x}\,)\cdot\hat{n}}{aH}\hat{n} \right),
	$}
\end{eqnarray}
where the subscript ``${\rm s}$'' denotes a quantity defined in redshift space. Then, $P_{\rm pv}$ and $\xi_{\rm pv}$ are represented as
\begin{eqnarray}
	\scalebox{0.95}{$\displaystyle
  \hspace{-0.5cm}(2\pi)^3 \delta_{\rm D}( \vec{k}+\vec{k}' ) P_{\rm pv}(\vec{k}) 
	$}\hspace{-0.25cm}
	&=& \hspace{-0.25cm}
	\scalebox{0.95}{$\displaystyle
	\left\langle p_{\rm s}(\vec{k})\,  \delta_{\rm s}(\vec{k}') - \delta_{\rm s}(\vec{k})\, p_{\rm s}(\vec{k}') \right\rangle $}
	\nonumber \\
	\scalebox{0.95}{$\displaystyle
	\xi_{\rm pv}(\vec{s}_1-\vec{s}_2) 
	$}\hspace{-0.25cm}
	&=&\hspace{-0.25cm}
	\scalebox{0.95}{$\displaystyle
	\Big\langle p_{\rm s}(\vec{s}_1)\, \delta_{\rm s}(\vec{s}_2) - \delta_{\rm s}(\vec{s}_1)\, p_{\rm s}(\vec{s}_2) \Big\rangle, $}
\end{eqnarray}
where $\delta_{\rm s}$ is the redshift-space density contrast.

In our recent studies~\citep{Sugiyama:2015dsa,Sugiyama2016arXiv160606367S}, we have shown that the LOS pairwise velocity power spectrum including the RSD effect is related to the galaxy power spectrum in redshift space $P_{\rm s}(\vec{k})$ under the global plane parallel approximation $\hat{n}_i\sim\hat{n}_j\sim\hat{n}$,
\footnote{
While equation~(\ref{Eq:pair0}) represents the redshift-space $P_{\rm pv}$, 
the real-space one without including the RSD effect can be given by taking $\gamma = 0$ in equation~(\ref{Eq:pair0})
(see also~\cite{Scoccimarro2004PhRvD..70h3007S}):
\begin{eqnarray}
	P_{\rm pv}(\vec{k}) = \left( i\frac{aH}{\vec{k}\cdot\hat{n}} \right) 
	\frac{\partial}{\partial \gamma}P_{\rm s}(\vec{k};\gamma\, \vec{v}\,)\bigg|_{\gamma = 0}. \nonumber
\end{eqnarray}
}
\begin{eqnarray}
	P_{\rm pv}(\vec{k}) = \left( i\frac{aH}{\vec{k}\cdot\hat{n}} \right) 
	\frac{\partial}{\partial \gamma}P_{\rm s}(\vec{k};\gamma\, \vec{v}\,)\bigg|_{\gamma = 1},
	\label{Eq:pair0}
\end{eqnarray}
where the dependence of $P_{\rm s}(\vec{k})$ on peculiar velocity that causes RSDs is schematically represented as $P_{\rm s}(\vec{k};\vec{v}\,)$. This relation can be easily shown by substituting the particle description of the galaxy power spectrum
\begin{eqnarray}
	  P_{\rm s}(\vec{k}) =  \left\langle \frac{V}{N^2} \sum_{i,j} 
	  e^{-i\vec{k}\cdot\vec{s}_{ij}} \right\rangle
\end{eqnarray}
into equation~(\ref{Eq:pair0}), where $\vec{s}_i$ is the redshift-space coordinates (equation~\ref{x_to_s}).

It is worth noting that the relation in equation~(\ref{Eq:pair0}) is quite general and holds without any assumption except for the global plane parallel approximation. If dark matter obeys the system of equations for a pressure-less perfect fluid without vorticity, and if $f = \Omega_{\rm m}^{1/2}$, where $f$ denotes the linear growth rate, we can solve the equations using cosmological perturbation theory. These assumptions lead to separable solutions of redshift $z$ (or the linear growth function $D(z)$) and wavevector $\vec{k}$ to arbitrary order in perturbation theory, resulting in that the perturbation theory solution of the velocity field is proportional to the linear growth rate to any order: $\vec{v} \propto f$ (or $f\sigma_8$)~\citep{Bernardeau2002PhR...367....1B}. Since the approximation that $f\approx \Omega_{\rm m}^{0.557-0.02z}$ is accurate to $0.3\%$~\citep{Polarski2008PhLB..660..439P}, the separable solutions in non-linear perturbation theories have been widely used. Then, equation~(\ref{Eq:pair0}) is simplified to
\begin{eqnarray}
	  P_{\rm pv}(\vec{k}) = \left( i\frac{aHf}{\vec{k}\cdot\hat{n}} \right) \frac{\partial}{\partial f}\,P_{\rm s}(\vec{k};\vec{v}\,).
	\label{Eq:pair}
\end{eqnarray}
The above relation holds even if the linear growth rate $f$ is replaced by $f\sigma_8$. These relations in equations~(\ref{Eq:pair0}) and (\ref{Eq:pair}) can be applied to any object such as dark matter particles, halos, galaxies, and galaxy clusters; furthermore, they could be used to compute the LOS pairwise velocity power spectrum when we use any cosmological perturbation theory: e.g., the standard perturbation theory~\citep{Bernardeau2002PhR...367....1B}, the Lagrangian perturbation theory~\citep{Matsubara2008PhRvD..77f3530M,Carlson2013MNRAS.429.1674C,Sugiyama2014ApJ...788...63S}, the effective field theory approach~\citep{Baumann2012JCAP...07..051B,Carrasco2012JHEP...09..082C}, the distribution function approach~\citep{Seljak2011JCAP...11..039S}, and the TNS model~\citep{Taruya2010PhRvD..82f3522T}.

\begin{figure*}
	\includegraphics[width=1.0\textwidth]{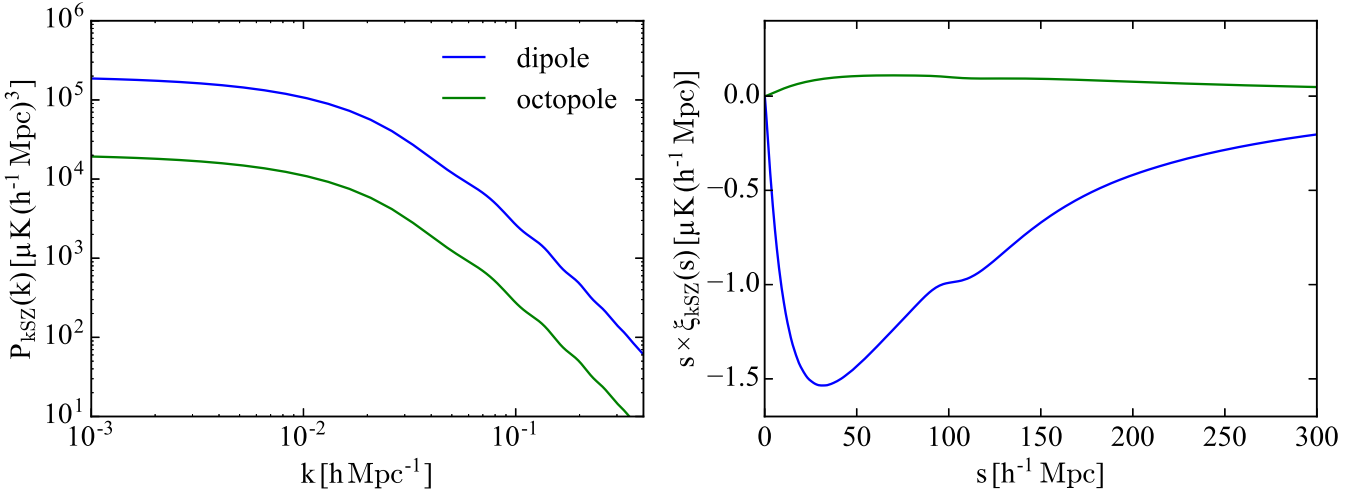}
	\caption{
Dipoles ($\ell=1$) and octopoles ($\ell=3$) of both the pairwise kSZ power spectrum (left panel) and correlation function (right panel) for the LOWZ sample. We adopt the fiducial values of $f\sigma_8=0.392$ and $b\sigma_8=1.283$ and the best-fitting value of the optical depth obtained from our kSZ analysis (Sec.~\ref{Sec:ResultsPairwisePower}), $\tau E = 3.95\times10^{-5}$. The dipole is about $10$ times larger than the octopole.
}
	\label{fig:theory_pkxi}
\end{figure*}

We compute the LOS pairwise velocity power spectrum in linear perturbation theory~\citep{Okumura2014JCAP...05..003O,Sugiyama:2015dsa}. Using equation~(\ref{Eq:pair}), we derive
\begin{eqnarray}
	  P_{\rm pv}(\vec{k}) &=&
	  \left( i\frac{aHf}{\vec{k}\cdot\hat{n}} \right) \frac{\partial}{\partial f} \left( b + f \mu^2 \right)^2 D^2 P_{\rm lin}(k) \nonumber \\
	&=& 2iaHf\mu\left( b + f\mu^2 \right) D^2\frac{P_{\rm lin}(k)}{k},
	\label{pk_lin}
\end{eqnarray}
where $\left( b + f\mu^2 \right)$ is the linear Kaiser factor~\citep{Kaiser1987MNRAS.227....1K}, $\mu=\hat{k}\cdot\hat{n}$, $b$ is the linear bias, and $P_{\rm lin}$ is the linear matter power spectrum at the present time. Similarly to the RSD analysis of the galaxy power spectrum (as the latest result, see \cite{Alam2016arXiv160703155A} and references therein), we expand $P_{\rm pv}$ in Legendre polynomials ${\cal L}_{\ell}$,
\begin{eqnarray}
	  P_{ {\rm pv},\, \ell}(k) = (2\ell+1) \int \frac{d\Omega_k}{4\pi}\, {\cal L}_{\ell}(\mu) P_{\rm pv}(\vec{k}),
	\label{Legendre}
\end{eqnarray}
where $d\Omega_k$ is the solid angle element in $k$-space. 
In the linear theory limit, the pairwise kSZ power multipoles become zero except for $\ell=1$ and $3$, and using equations~(\ref{kSZ_pk}), (\ref{pk_lin}), and (\ref{Legendre}), we finally have,
\begin{eqnarray}
	\scalebox{0.85}{$\displaystyle
	P_{ {\rm kSZ},\, \ell=1}(k) $}
	&=& 
	\scalebox{0.85}{$\displaystyle
	2i\frac{T_0 H_0}{c}(a\tau E)
	\left[ (f\sigma_8) (b \sigma_8) + \frac{3}{5}\left( f\sigma_8 \right)^2 \right] \frac{P_{\rm lin}(k)}{k} $}
	\nonumber \\
	\scalebox{0.85}{$\displaystyle
	P_{ {\rm kSZ},\, \ell=3}(k) $}
	&=&
	\scalebox{0.85}{$\displaystyle
	2i\frac{T_0 H_0}{c}(a\tau E) \left[ \frac{2}{5}\left( f\sigma_8 \right)^2 \right] \frac{P_{\rm lin}(k)}{k},$}
	\label{Eq:model_pk}
\end{eqnarray}
where $E(z) = H(z)/H_0$, which is given by $E(z) = \sqrt{(1+z)^3\Omega_{\rm m} + \Omega_{\rm \Lambda}}$ in a flat $\Lambda$CDM model, and we normalize the linear power spectrum using $\sigma_8$ at the present time: $P_{\rm lin}(k) \to P_{\rm lin}(k)/\sigma^2_8(z=0)$. It is worth noting that the pairwise kSZ power constrains $\tau E$, not $\tau$, where the growth rate $f \sigma_8$ and the linear bias $b\sigma_8$ are well constrained from the RSD analysis of galaxy clustering: $f\sigma_8=0.392\, (0.445)$ and $b\sigma_8 = 1.283\, (1.218)$ for the LOWZ (CMASS) sample~\citep{Gil-Marn2016MNRAS.460.4188G}. In contrast to the galaxy power spectrum that is real and contains only even multipoles, the pairwise kSZ power spectrum is imaginary and contains only odd multipoles.

The inverse Fourier transform of the pairwise kSZ power multipoles leads to the analytic form of the pairwise kSZ correlation function
\begin{eqnarray}
	\xi_{ {\rm kSZ},\, \ell}(s) = i^{\ell}\int \frac{dk\,k^2}{2\pi^2}j_{\ell}(ks)P_{ {\rm kSZ},\, \ell}(k),
\end{eqnarray}
where $j_{\ell}(ks)$ is the spherical Bessel function of order $\ell$. In linear theory, we obtain from equation~(\ref{Eq:model_pk}) 
\begin{eqnarray}
	\scalebox{0.95}{$\displaystyle
	\xi_{ {\rm kSZ},\, \ell=1}(s) $}
	\hspace{-0.25cm} &=&\hspace{-0.25cm}
	\scalebox{0.95}{$\displaystyle
	 - 2\frac{T_0 H_0}{c}\left( a\tau E\right)
	 \left[ (f\sigma_8) (b \sigma_8) + \frac{3}{5}\left( f\sigma_8 \right)^2 \right] \xi'_{\ell=1}(s) $}
	\nonumber \\
	\scalebox{0.95}{$\displaystyle
	\xi_{{\rm kSZ},\, \ell=3}(s) $}
	\hspace{-0.25cm}&=&\hspace{-0.25cm}
	\scalebox{0.95}{$\displaystyle
	2\frac{T_0H_0}{c}\left( a\tau E \right)
	\left[ \frac{2}{5}\left( f\sigma_8 \right)^2 \right] \xi'_{\ell=3}(s), $}
	\label{Eq:model_xi}
\end{eqnarray}
where
\begin{eqnarray}
	  \xi'_{\ell}(s) =  \int \frac{dk\, k^2}{2\pi^2}j_{\ell}(ks)\frac{P_{\rm lin}(k)}{k}. 
\end{eqnarray}
We find that the pairwise kSZ dipole, $\xi_{ {\rm kSZ},\, \ell=1}$, is negative in linear theory. This means that gravitational attraction predicts a slight tendency of any pair of galaxies to be moving toward rather than away from each other at large scales, where we follow the convention that if two galaxies are moving toward each other, their contribution to the pairwise velocity is negative. The linear Kaiser effect from RSDs increases the amplitude of the pairwise kSZ dipole by $\sim 20\%$ for both LOWZ and CMASS at linear regimes. On the other hand, at non-linear small scales RSDs displace galaxies away from each other, and thus, in redshift space the sign of $\xi_{ {\rm kSZ},\, \ell=1}$ changes around $10\hMpc$ ($k\simeq0.2\hk$) from negative to positive~\citep{Okumura2014JCAP...05..003O,Sugiyama:2015dsa,Sugiyama2016arXiv160606367S}. This small-scale effect of RSDs is analogous to the Finger-of-God (FoG) effect but is not the same, because we have shown that this change in sign of $\xi_{ {\rm kSZ},\, \ell=1}$ appears even in the Zel'dovich approximation that does not form halos~\citep{Sugiyama:2015dsa}. Therefore, it could be considered that the change in sign of $\xi_{ {\rm kSZ},\, \ell=1}$ is caused by non-linear velocity dispersion effects, part of which is the FoG effect. The latest kSZ measurement using the high resolution ACT data~\citep{DeBernardis2016arXiv160702139D} clearly shows the expected change in sign of the pairwise velocity at $\sim 10\hMpc$.

Fig.~\ref{fig:theory_pkxi} shows the theoretical predictions of the pairwise kSZ dipole and octopole in linear theory (equations~\ref{Eq:model_pk} and~\ref{Eq:model_xi}) for the LOWZ sample ($f\sigma_8=0.392$ and $b\sigma_8=1.283$), where we use the best fit value of the optical depth, $\tau E = 3.95\times10^{-5}$, estimated from our kSZ measurement (Sec.~\ref{Sec:ResultsPairwisePower}). 
Since this figure shows the linear theory results that we use in our analysis,
the change in sign of $\xi_{\rm kSZ, \ell=1}$ around $\sim 10\hMpc$ does not occur 
unlike non-linear theories~\citep{Okumura2014JCAP...05..003O,Sugiyama:2015dsa}.
As the shape of the pairwise kSZ power spectrum is proportional to $P_{\rm lin}(k)/k$, the amplitude increases at large scales. As the data improves, we will be able to investigate the pairwise kSZ signal at larger scales than $\sim 200\hMpc$~\citep[e.g.][Fig. 3]{DeBernardis2016arXiv160702139D}.  While the data should show a  baryon acoustic peak around $\sim 100\hMpc$ in the linear theory prediction of $\xi_{ {\rm kSZ},\, \ell=1}$ (see blue line in the right panel of Fig.~\ref{fig:theory_pkxi}), the peak is significantly broadened by non-linear effects~\citep{Sugiyama:2015dsa}. 
The octopole, which is generated from the RSD effect and is free of the linear bias, is about $10$ times smaller than the dipole. In future kSZ measurements such as Advanced ACTPol and CMB-S4, we will want to measure all of the multipoles up to the octopole. This full measurement will constrain cosmological parameters, especially for the linear growth rate $f \sigma_8$ and the Hubble expansion parameter $H$~\citep{Sugiyama2016arXiv160606367S}. However, the data used in this work is too noisy to detect the octopole; therefore, we do not analyze it in what follows. In this paper, we measure the pairwise kSZ power dipole $P_{ {\rm kSZ},\, \ell=1}$ and constrain the optical depth value $\tau E$ after fixing all the other parameters.

\subsection{Comparison with previous works}

It is well known that the commonly used pairwise velocity $v_{\rm pair}(r)$ is related to the time derivative of the galaxy two-point correlation function $\xi_{\rm g}(r)$ under the following three assumptions: pair conservation of galaxies, isotropic peculiar velocities, and isotropic clustering of galaxies~\citep{Peebles1976Ap&SS..45....3P,Davis1977,Peebles1980lssu.book.....P},
\begin{eqnarray}
	  v_{\rm pair}(r) = - \frac{\int_0^r dr'\, r'^2\, \frac{\partial}{\partial t}\,\xi_{\rm g}(r')}{r^2\,(1+\xi_{\rm g}(r))}.
	  \label{pairwise_v}
\end{eqnarray}
Since the total number of galaxies is not conserved, and since we measure the LOS velocity via the kSZ effect in anisotropic galaxy clustering by RSDs, it will not be possible to apply the above expression to the kSZ analysis directly. We stress here that the relations presented in equations~(\ref{Eq:pair0}) and (\ref{Eq:pair}) hold without any of these three assumptions.

The dipole of the pairwise kSZ correlation function $\xi_{ {\rm kSZ},\, \ell=1}$ is closely related to the mean pairwise velocity $v_{\rm pair}$ (equation~\ref{pairwise_v}) estimated from the previous kSZ measurements~\citep{Hand2012,Planck2016A&A...586A.140P,Soergel2016MNRAS.461.3172S,DeBernardis2016arXiv160702139D}. The pairwise velocity $v_{\rm pair}$ is computed by the pair-weighted average with the weight factor $n_1 n_2 / \langle n_1 n_2\rangle$, where $n_1$ and $n_2$ are the number density fields at points $\vec{x}_1$ and $\vec{x}_2$, and $\langle n_1 n_2\rangle$ yields the galaxy two-point correlation function $\xi_{\rm g}$. On the other hand, we replace the denominator of the pair-weighted average $\langle n_1 n_2\rangle$ by the square of the mean number density $\bar{n}^2$. Thus, our approach does not need an additional calculation of $\xi_{\rm g}$ in modeling $\xi_{ {\rm kSZ},\, \ell}$ (see \cite{Okumura2014JCAP...05..003O} for the detailed discussion of the difference between pair-weighted and density-weighted velocities).

\citet{Keisler2013} noted  that the kSZ signal is proportional to $\tau b f \sigma_8^2$ and can be used to break the degeneracy between $f$ and $\sigma_8$ that appears in the RSD analysis of galaxy clustering. On the other hand, in our analysis the pairwise kSZ dipole is proportional to $\tau \left[ (b\sigma_8)(f\sigma_8) + \frac{3}{5}(f\sigma_8)^2 \right]$ due to the RSD effect, and we fix $f \sigma_8$ and $b\sigma_8$ from the RSD analysis; therefore, the degeneracy between $f$ and $\sigma_8$ cannot be broken.

\section{METHODOLOGY}
\label{Sec:Methodology}

\subsection{Galaxy and kSZ density fluctuations}

We estimate the filtered CMB temperature $T_{\rm AP}(\hat{n}_i)$ associated with galaxy $i$ by applying the AP filter (Sec.~\ref{Sec:kSZ}).
Using HEALPIX, we firstly find pixels that lie within a disk of radius, $\theta_{\rm c}$, around each galaxy and average the temperature within the disk.
Next, we compute the mean temperature in a surrounding ring of inner and outer radii and subtract it from the mean temperature within the disk,
where the outer radius is chosen to be $\sqrt{2} \theta_{\rm c}$.
Finally, we define the filtered CMB temperature fluctuation $\delta T_{\rm AP}(\hat{n}_i)$ by subtracting the average of $T_{\rm AP}(\hat{n}_i)$ obtained from all galaxies after weighting by $w_{\rm c}$ from $T_{\rm AP}(\hat{n}_i)$:
\begin{eqnarray}
	\delta T_{\rm AP}(\hat{n}_i) = T_{\rm AP}(\hat{n}_i) 
	- \frac{\sum_i^{N_{\rm gal}} T_{\rm AP}(\hat{n}_i)w_{ {\rm c}}(\vec{s}_i)}{\sum_i^{N_{\rm gal}} w_{\rm c}(\vec{s}_i)}.
	\label{Eq:DeltaT_AP}
\end{eqnarray}
where $w_{\rm c}$ is the completeness weight defined in equation~(\ref{Weight}).

The fluctuations of three-dimensional galaxy and kSZ density fields are then given by
\begin{eqnarray}
	  \delta n(\vec{s}\,) &=&  \sum_i^{N_{\rm gal}}  w_{ {\rm c}}(\vec{s}_i) \delta_{\rm D}\left( \vec{s} - \vec{s}_i \right)
	  - \alpha \sum_i^{N_{\rm ran}}  \delta_{\rm D}\left( \vec{s} - \vec{s}_i \right), \nonumber \\
	  \delta T(\vec{s}\,) &=&  \sum_i^{N_{\rm gal}}
	  \delta T_{ {\rm AP}}(\hat{n}_i) w_{ {\rm c}}(\vec{s}_i) \delta_{\rm D}\left( \vec{s} - \vec{s}_i \right),
	  \label{Eq:fluctuation}
\end{eqnarray}
where $N_{\rm gal}$ and $N_{\rm ran}$ represent the numbers of galaxies in the real and synthetic catalogues, and the factor $\alpha$ is the ratio between the weighted numbers of galaxies in the galaxy and random catalogues, $\alpha = \sum_i^{N_{\rm gal}}w_c(\vec{s}_i) /N_{\rm ran}\sim0.01$. Here we construct the fluctuations $\delta n$ and $\delta T$ so that their volume-averages become zero: $\int d^3s\, \delta T(\vec{s}\,) =  \int d^3s\, \delta n(\vec{s}\,) = 0$.

Any evolution of contaminations of the kSZ signal, e.g. the tSZ effect and cosmic infrared background (CIB) emission, with redshift would result in a bias effect. Therefore, almost all the previous works compute the mean measured temperature weighing by a Gaussian kernel as a function of redshift, $G(z_i,z_j,\sigma_z) = \exp\left[ -\left( z_i - z_j \right)^2/(2\sigma_z^2) \right]$, and subtract it from the AP filtered temperature $T_{\rm AP}$~\citep{Hand2012,Planck2016A&A...586A.140P,Soergel2016MNRAS.461.3172S,DeBernardis2016arXiv160702139D} to obtain an unbiased estimate of the kSZ signal. However, since there is minimal CIB contamination in the WPR2 maps~\citep{Bobin:2015fgb}, we do not adopt this prescription. To validate our analysis, we compute the pairwise kSZ power spectrum using this prescription with $\sigma_{z}=0.02$ and compare it to that without using the prescription. We have then checked that any residual CIB contamination is subdominant in our analysis and produces a small change in the pairwise kSZ power spectrum estimates: the relative difference is $\leq 5\%$ at the scales of interest ($k<0.15\hk$).

\subsection{Estimator of the Pairwise kSZ power multipoles}
\label{Sec:Estimator}

In analogy to the estimator of the galaxy power multipoles~\citep{Feldman1994ApJ...426...23F,Yamamoto2006PASJ...58...93Y}, we present an estimator of the pairwise kSZ power multipoles as,
\begin{eqnarray}
	  \widehat{P}_{\rm kSZ, \ell}(\vec{k})
	&=&  - \frac{\left( 2\ell+1 \right)}{A}
	\int d^3s_1 \int d^3s_2\, e^{-i\vec{k}\cdot\vec{s}_{12}}
	{\cal L}_{\ell}(\hat{k}\cdot\hat{n}_{12}) \nonumber \\
	&\times& \Big[ \delta T(\vec{s}_1) \delta n(\vec{s}_2) - \delta n(\vec{s}_1)\delta T(\vec{s}_2) \Big],
	\label{Eq:estimator1}
\end{eqnarray}
where $\vec{s}_{12}=\vec{s}_1-\vec{s}_2$ is the relative coordinates between two different points $\vec{s}_1$ and $\vec{s}_2$, and the unit vector $\hat{n}_{12}$ of $\vec{n}_{12} = (\vec{s}_1+\vec{s}_2)/2$ is used as the LOS direction to the pair of points $\vec{s}_1$ and $\vec{s}_2$. The normalization $A$ is given by
\begin{eqnarray}
	A =  \int d^3s\, \bar{n}^2(\vec{s}\,),
	\label{Eq:A}
\end{eqnarray}
where $\bar{n}$ is the number density measured from the random catalogue, multiplied by $\alpha$,
\begin{eqnarray}
	  \bar{n}(\vec{s}\,) = \alpha \sum_i^{N_{\rm ran}}\delta_{\rm D}\left( \vec{s} - \vec{s}_i \right).
	  \label{Eq:n_bar}
\end{eqnarray}

The utility of the \textit{pairwise estimator} is apparent in Fourier space, as the difference in the square bracket of equation~(\ref{Eq:estimator1}), $\delta T(\vec{s}_1) \delta n(\vec{s}_2) - \delta n(\vec{s}_1)\delta T(\vec{s}_2)$, can be used  to extract the imaginary elements of the pairwise kSZ power multipoles. We note here that the shot noise term does not appear in the pairwise estimator, as it is canceled out by the difference.

There are two main sources of noise in the pairwise kSZ signal, detector noise and primary CMB anisotropies. Integrating over $\vec{s}_1$ and $\vec{s}_2$ after weighting by odd-pole Legendre polynomials in the pairwise estimator (equation~\ref{Eq:estimator1}), these noise components will average to zero, because the parity of them is even, while the kSZ signal is odd parity. In other words, the noise contribution will not bias the imaginary part of the pairwise kSZ power multipoles, because they are scalar quantities, not a LOS component of the peculiar velocity vector such as the kSZ signal. While the noise does not contribute to the expectation value for the signal, it does contribute to its variance as we discuss in  Appendix~\ref{Ap:Covariance}.

In this paper, we apply the local plane parallel approximation:
\begin{eqnarray}
	{\cal L}_{\ell}(\hat{k}\cdot\hat{n}_{12}) \approx {\cal L}_{\ell}(\hat{k}\cdot\hat{s}_1)   \approx{\cal L}_{\ell}(\hat{k}\cdot\hat{s}_2).
	\label{Eq:parallel}
\end{eqnarray}
Then, this approximation allows the integrals in equation~(\ref{Eq:estimator1}) to decouple into a product of Fourier transforms~\citep{Yamamoto2006PASJ...58...93Y,Blake2011MNRAS.415.2876B,Beutler2014MNRAS.443.1065B,Samushia2015MNRAS.452.3704S}:
\begin{eqnarray}
	  \widehat{P}_{\rm kSZ, \ell}(\VEC{k})
	  =  - \frac{\left( 2\ell+1 \right)}{A} \left[   \delta T_{\ell}(\VEC{k})\, \delta n^*(\VEC{k}) - \mathrm{c.c.} \right],
	\label{Eq:estimator2}
\end{eqnarray}
where
\begin{eqnarray}
	\delta n(\VEC{k}) &=&  \int d^3s\, \mathrm{e}^{-\mathrm{i}\VEC{k}\cdot\VEC{s}}\,  \delta n(\VEC{s}) \nonumber \\
	\delta T_{\ell}(\VEC{k}) &=&  \int d^3s\, \mathrm{e}^{-\mathrm{i}\VEC{k}\cdot\VEC{s}}\, {\cal L}_{\ell}(\hat{k}\cdot\hat{s})\, \delta T(\VEC{s}).
	\label{Eq:F_l}
\end{eqnarray}
We note here that $\delta n(\VEC{k})$ and $\delta T_{\ell}(\VEC{k})$ can be built to be computable using any fast Fourier transform (FFT) algorithm
~\citep{Bianchi2015MNRAS.453L..11B,Scoccimarro2015PhRvD..92h3532S}:
therefore, the computation of the estimator will be ${\cal O}(N_k \log N_k)$, where $N_k$ is the numbers of $k$-modes 
(for details, see Appendix~\ref{Ap:Estimator}).

Finally, we average the amplitude of the pairwise kSZ power multipoles measured from equation~(\ref{Eq:estimator2}) in spherical shell of $\vec{k}$ to produce our estimate of $\widehat{P}_{ {\rm kSZ},\ell}(k)$ at each bin $k=|\vec{k}|$:
\begin{eqnarray}
	\widehat{P}_{ {\rm kSZ}, \ell}(k) &=& \int \frac{d\Omega_k}{4\pi}  \widehat{P}_{ {\rm kSZ}, \ell}(\vec{k}) \nonumber \\
	&=& \frac{1}{N_{\rm mode}} \sum_{k - \frac{\Delta k}{2}< k < k + \frac{\Delta k}{2}} \widehat{P}_{ {\rm kSZ}, \ell}(\vec{k}),
	\label{Eq:avPk}
\end{eqnarray}
where $N_{\rm mode}$ is the number of independent Fourier modes in a $k$-bin.
\subsection{Prescription of Measurements}

\citet{Gil-Marn2016MNRAS.460.4188G} computed the galaxy power spectra at the NGC and SGC and combined them by weighting by their effective area. Since we use the values of $f\sigma_8$ and $b\sigma_8$ measured from the RSD analysis of galaxy clustering in~\citet{Gil-Marn2016MNRAS.460.4188G}, for consistency we compute the combined pairwise kSZ power multipoles 
\begin{eqnarray}
	\scalebox{0.85}{$\displaystyle
	P_{{\rm kSZ}, \ell}(k) = 
	\left( A_{ {\rm NGC} } P_{ {\rm kSZ}, \ell}^{(\rm NGC)} + A_{\rm SGC} P_{ {\rm kSZ}, \ell}^{(\rm SGC)} \right)
	/  \left( A_{\rm SGC} + A_{\rm NGC} \right) ,$}
	\label{Eq:NGC_SGC}
\end{eqnarray}
where $A_{\rm NGC}$ and $A_{\rm SGC}$ are the effective areas of the NGC and SGC, respectively, whose values are $A_{\rm NGC}^{\rm LOWZ} = 5836\, {\rm deg}^2$, $A_{\rm SGC}^{\rm LOWZ} = 2501\,{\rm deg}^2$, $A_{\rm NGC}n^{\rm CMASS} = 6851\, {\rm deg}^2$, and $A_{\rm SGC}^{\rm CMASS} = 2525\, {\rm deg}^2$. In the following analysis we perform the parameter fitting process to the combined NGC+SGC power spectrum.

We define the Cartesian coordinates $\vec{s}=\left( x,y,z \right)$ with $z$ being the axis toward the north pole. We place the LOWZ and CMASS samples in a cuboid of dimensions $\left( L_x,\, L_y,\, L_z \right)$, where $\left( L_x,\, L_y,\, L_z \right)\, [\hMpc]$ $ = \left( 2400,\, 4200,\, 2400 \right)$ for CMASS-NGC, $\left( 2600,\, 3400,\, 2000 \right)$ for CMASS-SGC, $\left( 2300,\, 4000,\, 2300 \right)$ for LOWZ-NGC, and $\left( 2400,\, 3200,\, 1700 \right)$ for LOWZ-SGC. The CMASS and LOWZ galaxies are distributed on the FFT grid using the triangular-shaped cloud (TSC) assignment function. We use the Fast Fourier Transform in the West (FFTW)\footnote{ \url{http://fftw.org} } with a $512$ grid on each axis. This corresponds to a grid-cell resolution of $\sim 5\hMpc$ for both CMASS and LOWZ. We have checked that for the scales of interest ( $k < 0.15\hk$), doubling the number of grid-cells per side, from $512$ to $1024$, produces a small change in the pairwise kSZ power spectrum estimated by equation~(\ref{Eq:estimator2}), within $0.3\%$.

The FFT algorithm requires the interpolation of functions on a regular grid in position space. This interpolation results in the so-called aliasing effect (e.g.~\citet{Jing2005ApJ...620..559J}). To reduce the aliasing effect, we adopt a simple technique proposed by~\citep{HockneyEastwood1981,Sefusatti2016MNRAS.460.3624S} that is based on the interlacing of two grids (for details, see Appendix~\ref{Sec:PrescriptionOfFFTs}). 

In our analysis, we do not take account of the binning discreteness. It is evident at low $k$ and may introduce systematic biases into the kSZ power estimate at a few percent level. We defer a more detailed analysis investigating this effect for future kSZ measurements.

\section{SURVEY WINDOW FUNCTIONS}
\label{Sec:SurveyWindowFunction}

\begin{figure}
	\includegraphics[width=\columnwidth]{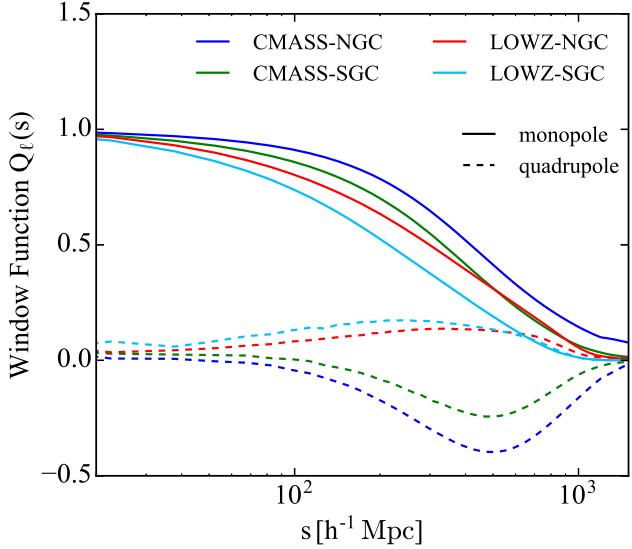}
	\caption{
Multipoles of survey window function for BOSS DR12 as given in equation~(\ref{Eq:Q2}), which are used to compute the masked pairwise kSZ power dipole in equation~(\ref{Eq:Template}). The coloured lines display the window functions for CMASS-NGC (blue), CMASS-SGC (green), LOWZ-NGC (red), and LOWZ-SGC (cyan). The solid and dashed lines denote the monopole and quadrupole, respectively. The monopole effect decreases with increasing survey volumes, so that the monopole $Q_0$ becomes close to unity. The quadrupole $Q_2$ represents survey geometry anisotropies along radial directions, and its shape depends on a given survey: e.g. the quadrupole is positive for LOWZ and is negative for CMASS.
	}
	\label{fig:window_function}
\end{figure}

\begin{figure}
	\includegraphics[width=\columnwidth]{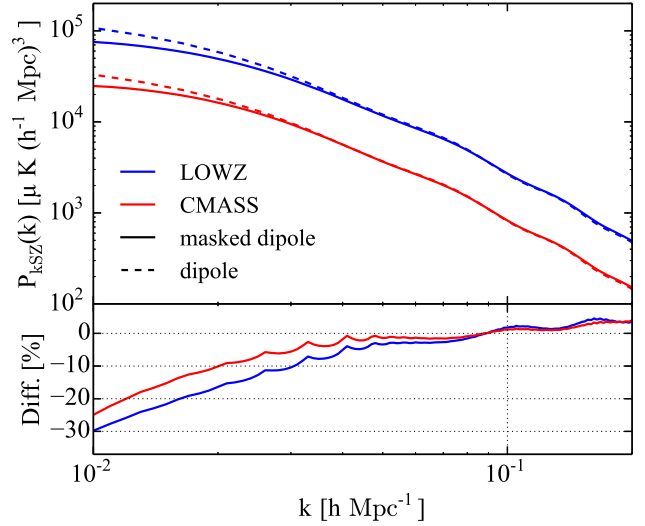}
	\caption{
Theoretical predictions of the pairwise kSZ power dipole. The blue and red lines represent the results of LOWZ and CMASS, respectively.
While the dashed lines show the true pairwise kSZ power dipole, the solid lines denote the masked results.
Bottom panels show the relative differences: $P_{\rm masked}/P_{\rm true}-1$. The window function effects are of order of $20\mathchar`-30\%$ around $k=0.01\hk$.
	 }
	\label{fig:theory_pkxi_window}
\end{figure}

The imprint of survey geometry yields systematic differences between the observed pairwise kSZ power spectrum and their theoretical predictions presented in Sec.~\ref{Sec:kSZ_Power}. We follow the treatment of survey window functions suggested by~\citet{Wilson2015arXiv151107799W} (see also~\citet{Beutler2016arXiv160703150B}) and apply it to our kSZ analysis. While the survey window effect in the pairwise kSZ power spectrum is represented by a convolution integral in Fourier space, from the convolution theorem, the pairwise kSZ correlation function and the window function multiply to give the masked correlation function in configuration space. Therefore, we compute the masked pairwise kSZ power spectrum through a Hankel transform of the masked pairwise kSZ correlation function. In this section, we present the main equations, leaving their full derivations in Appendix~\ref{Ap:Derivations}.

To characterize the distortion of the survey geometry,
we define the survey window function multipoles in configuration space:
\begin{eqnarray}
	Q_{\ell}(s) &=& \frac{\left( 2\ell+1 \right)}{A}
	\int \frac{d\Omega_s}{4\pi} \int d^3s_1 \int d^3s_2 \delta_{\rm D}\left( \vec{s} - \vec{s}_{12} \right) \nonumber \\
	&\times&{\cal L}_{\ell}(\hat{s}_{12}\cdot\hat{s}_{1})  \bar{n}(\vec{s}_1)\bar{n}(\vec{s}_2), 
	\label{Eq:Q1}
\end{eqnarray}
where we used the local plane parallel approximation, $\hat{s}_1\approx \hat{s}_2$,
and $\bar{n}(\vec{s})$ is the mean number density measured from the random catalogue (equation~\ref{Eq:n_bar}).

The masked (observed) pairwise kSZ power spectrum is expressed by the ensemble average of
its estimator (equation~\ref{Eq:estimator2}), namely $\langle \widehat{P}_{ {\rm kSZ},\, \ell}(k)\rangle$.
Using the window function multipoles, the masked pairwise kSZ power spectrum is given by
(for derivation, see Appendix~\ref{Ap:Derivations})
\begin{eqnarray}
	\left\langle \widehat{P}_{ {\rm kSZ},\, \ell}(k)\right\rangle
	&=&  4\pi (-i)^{\ell} (2\ell+1)\int ds\,s^2  j_{\ell}(ks) \nonumber \\
	&\times&\sum_{\ell_1\ell_2}  \left( \begin{smallmatrix} \ell & \ell_1 & \ell_2 \\ 0 & 0 & 0 \end{smallmatrix} \right)^2 
	\xi_{ {\rm kSZ},\, \ell_1}(s) Q_{\ell_2}(s),
	\label{Eq:xi_Q}
\end{eqnarray}
where the theoretical prediction of $\xi_{\rm kSZ,\, \ell}$ is given by equation~(\ref{Eq:model_xi}).
\footnote{It is worth noting that equation~(\ref{Eq:xi_Q}) reproduces equations~(19) and (20) in~\cite{Wilson2015arXiv151107799W} by replacing $\xi_{\rm kSZ}$ by the galaxy two-point correlation function $\xi_{\rm g}$.} For the pairwise kSZ power dipole, the necessary linear combination is obtained as 
\begin{eqnarray}
	\left\langle \widehat{P}_{ {\rm kSZ},\, \ell=1}(k)\right\rangle
	 \hspace{-0.25cm}&=& \hspace{-0.25cm}4\pi (-i)^{\ell} \int ds\,s^2  j_{\ell=1}(ks) \nonumber \\
	&\times& \hspace{-0.25cm} \xi_{ {\rm kSZ},\,  \ell=1}(s) \left(  Q_0(s) + \frac{2}{5} Q_2(s)  \right) ,
	\label{Eq:Template}
\end{eqnarray}
where in the right-hand-side the multipole expansion of the pairwise kSZ correlation function is truncated at the dipole, because higher multipoles (e.g. the octopole $\xi_{ {\rm kSZ},\,\ell=3}$) are so small as to be ignored. This equation is the final expression of our template model to fit to the measurement of the pairwise kSZ power dipole.

The multipole components of the survey window function used in our analysis are shown in Fig.~\ref{fig:window_function}. The monopole $Q_0$ should become zero at the scales larger than the survey volume ($\sim 1\, h^{-1}\,{\rm Gpc}$ for both CMASS and LOWZ), while on small scales where the survey edge effects no longer matter, it will be constant and will be equal to unity for an appropriate normalization $A$. Among the 4 regions of the BOSS, the correction derived from the monopole for CMASS-NGC is smallest while that for LOWZ-SGC largest. The quadrupole $Q_2$, which represents survey geometry anisotropies along radial directions, will average to zero on both large and small scales. Its shape significantly depends on a given survey geometry: the quadrupole becomes positive and negative for LOWZ and CMASS, respectively.

Fig.~\ref{fig:theory_pkxi_window} shows the theoretical predictions of the pairwise kSZ dipoles with the survey window corrections
for both LOWZ and CMASS. The amplitude of the masked $P_{ {\rm kSZ},\, \ell=1}$ becomes smaller than the true one at large scales, because we cannot find pairs of galaxies at scales larger than the survey volume. The window function effect is of order of $20\mathchar`-30\%$ around $k\sim0.01\hk$ and significantly increases for $k<0.01\hk$. 

\section{COVARIANCE ESTIMATES}
\label{Sec:Covariance}

We estimate the data covariance matrix through resampling of data itself, using the delete one jackknife method~\citep{Quenouille1956,Tukey1958}. After dividing the observed galaxy clustering data into $N_{\rm JK}$ spatial sub-regions on the sky, we define the resampling of the data by systematically omitting, in turn, each of $N_{\rm JK}$ sub-regions. To ensure that these regions cover approximately equal area, we employ the $k$-means algorithm\footnote{\url{https://github.com/esheldon/kmeans_radec/}}. To avoid underestimating clustering uncertainties, the resampling of the data should be applied on $N_{\rm JK}$ sub-regions into which the dataset has been split instead of on individual galaxies~\citep{Norberg2009MNRAS.396...19N}. Unless otherwise specified, we employ $N_{\rm JK}=200$.

In the jackknife scheme, the covariance matrix is estimated by
\begin{eqnarray}
	C_{ {\rm JK}, ij} =  \frac{N_{\rm JK}-1}{N_{\rm JK}} 
	  \sum_{k}^{N_{\rm JK}} \left( y^k_i - \bar{y}_i \right) \left( y^k_j - \bar{y}_j \right),
	  \label{Eq:cov_estimate}
\end{eqnarray}
where $y_i^k$ is the $i$th measure of the statistic of interest for the $k$th configuration: in our analysis, $y_i = \widehat{P}_{ {\rm kSZ},\, \ell=1}(k_i)$. It is assumed that the mean expectation value $\bar{y}_i$ is given by
\begin{eqnarray}
	\bar{y}_i = \frac{1}{N_{\rm JK}} \sum_k^{N_{\rm JK}} y_i^k.
\end{eqnarray}

The inverse of the covariance matrix $C_{\rm JK}^{-1}$ estimated from data itself using the jackknife method is a biased estimator. To correct for this bias, we rescale the inverse covariance matrix as~\citep{Hartlap2007A&A...464..399H}
\begin{eqnarray}
	C^{-1} = \frac{N_{\rm JK}-N_{\rm bin}-2}{N_{\rm JK}-1} C_{\rm JK}^{-1},
	\label{Eq:cov_correction}
\end{eqnarray}
where the Hartlap factor, $(N_{\rm JK}-N_{\rm bin}-2)/(N_{\rm JK}-1)$, accounts for the skewed nature of the inverse Wishart distribution, and $N_{\rm bin}$ is the number of bins. We use $7$ $k$-bins when measuring $\widehat{P}_{ {\rm kSZ},\, \ell=1}(k)$ (Sec.~\ref{Sec:ResultsPairwisePower}), resulting in a Hartlap factor of $0.96$.

In Appendix~\ref{Ap:Covariance}, we compare the variances (the diagonal entries of the covariance matrix) of the pairwise kSZ power dipole computed by linear theory developed in \citet{Sugiyama2016arXiv160606367S} with those estimated from data itself using the jackknife and bootstrap~\citep{Efron1979} methods for various spatial sub-regions on the sky. There, we find a good agreement between the theory and the data, implying the validities of both our variance model and the variance estimates from data itself.

\section{RESULTS}
\label{Sec:Results}

\begin{figure}
	\includegraphics[width=\columnwidth]{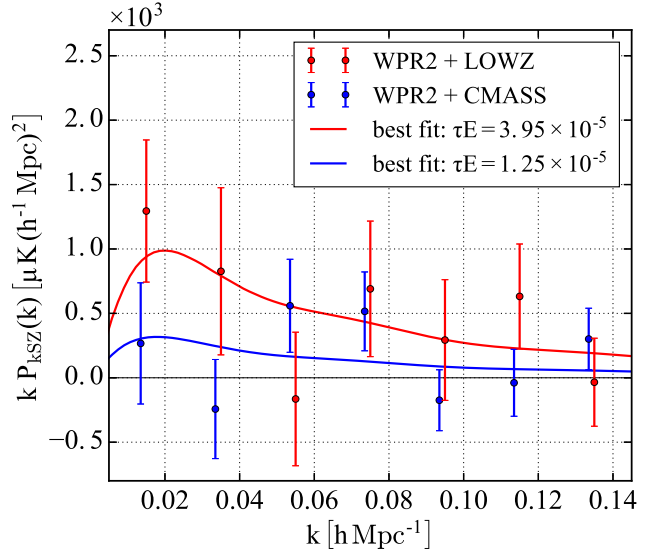}
	\caption{ 
Pairwise kSZ power spectrum dipoles measured from the WPR2 CMB map (Sec.~\ref{Sec:Data_CMB}) with both the LOWZ (red symbols) and CMASS (blue symbols) samples (Sec.~\ref{Sec:Data_Galaxy}) compared with the best fit models (Sec.~\ref{Sec:Theory}) including the window function correction (Sec.~\ref{Sec:SurveyWindowFunction}) (red and blue solid lines). The kSZ temperature per each galaxy is estimated using the AP filter with $5\arcmin\, (4\arcmin)$ aperture radius for the LOWZ (CMASS) sample. The errorbars shown are derived from the $1\sigma$ jackknife errors (Sec.~\ref{Sec:Covariance}). Using the fitting range $k=0.015\,\mathchar`-\,0.135\hk$ with $7$ separation bins, we find $\tau E = (3.95\pm1.62)\times10^{-5}$ for LOWZ and $\tau E= (1.25\pm1.06)\times10^{-5}$ for CMASS, corresponding to ${\rm S/N}=2.44$ and $1.18$, respectively.}
	\label{fig:pk}
\end{figure}

\begin{figure}
	\includegraphics[width=\columnwidth]{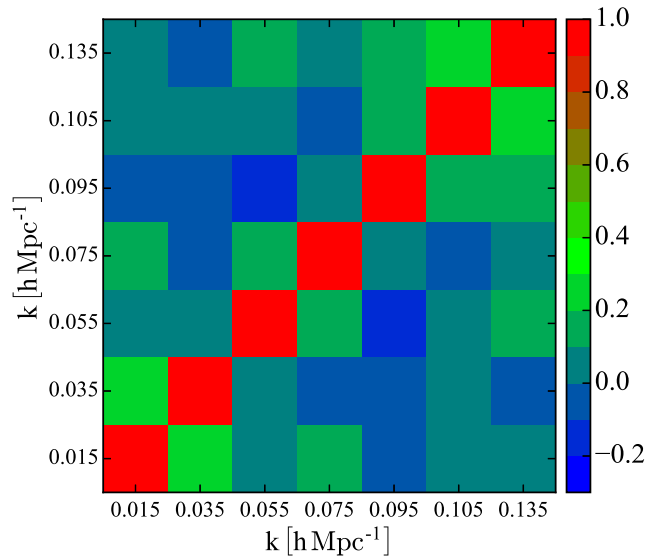}
	\caption{ 
Correlation coefficients of the pairwise kSZ power dipole measured from the WPR2 map with the LOWZ sample using the jackknife method. All the absolute values of the off-diagonal elements are less than $0.25$, showing weak correlation between different $k$-bins.
	}
	\label{fig:pk_covariance}
\end{figure}

\begin{figure}
	\includegraphics[width=\columnwidth]{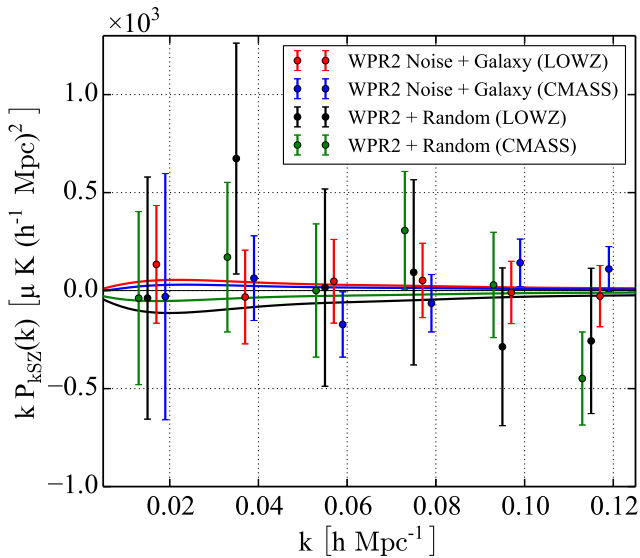}
	\caption{
		  Null tests in the following two ways for both the LOWZ and CMASS samples: the three-dimensional temperature fluctuation, $\delta T$ in equation~(\ref{Eq:fluctuation}), is estimated from (1) the WPR2 data map with the random catalogue and from (2) the WPR2 noise map with the galaxy data sample, where the galaxy density fluctuation, $\delta n$ in equation~(\ref{Eq:fluctuation}), is estimated from the galaxy data in all four cases. The pairwise kSZ power dipoles and their associated errors shown in this figure are estimated in the same way as Fig.~\ref{fig:pk}. The four solid lines are the best-fitting models for the null tests. In all four cases, the signals are consistent with null hypothesis.
	}
	\label{fig:pk_null}
\end{figure}

\begin{figure*}
	\includegraphics[width=1.0\textwidth]{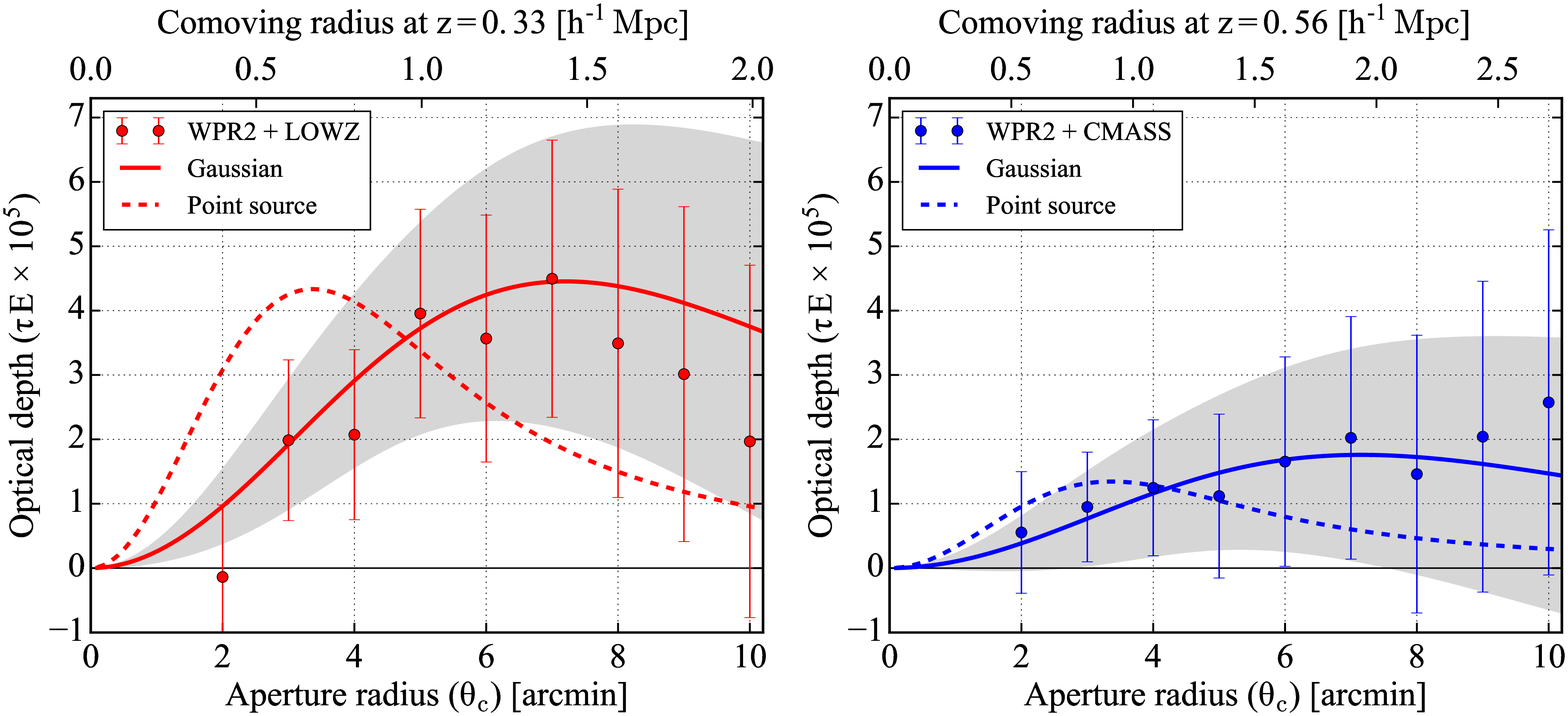}
	\caption{
		  Measured optical depth $\tau E$ as a function of the angular radius $\theta_{\rm c}$ of the AP filter using the WPR2 map with the LOWZ (left panel) and CMASS (right panel) galaxy samples. The $1\sigma$ error bars in this figure are derived from equation~(\ref{Eq:fitting}), and they are highly correlated (see Fig.~\ref{fig:tau_covariance}). We compare our measurements with the best fitting models assuming a projected Gaussian gas profile with a characteristic radius $\sigma_{\rm R}$, given by equations~(\ref{tau1}) and (\ref{Eq:gas_profile}) (solid lines). The grey shaded regions are the corresponding $1\sigma$ uncertainties computed by linear theory (equation~\ref{Eq:tau_cov_theory}). For the LOWZ (CMASS) sample, using the fitting range $\theta_{\rm c}=2\arcmin\,\mathchar`-\,10\arcmin\, (8\arcmin)$ with $1\arcmin$ bin width, the best fit parameter values are $f_{\rm gas}=0.188\, (0.236)$ and $\sigma_{\rm R}=4.02\, (3.98)\arcmin$. For a comparison, we plot the predictions from the point source model, which corresponds to $\sigma_{\rm R}=0$ in the Gaussian model, for $f_{\rm gas}=0.04$ (dashed lines). We find a clear shift outward in the peak of the AP filtered optical depth compared to the peak position predicted by the point source assumption. }
	\label{fig:tau_data_filter}
\end{figure*}

\subsection{Fitting prescription}

We fit the pairwise kSZ power dipole measured by its estimator in equation~(\ref{Eq:estimator2}) with the template given by equation~(\ref{Eq:Template}) with the optical depth $\tau E$ being a free parameter. In this paper, we do not analyze the pairwise kSZ correlation function, but we expect similar results to the analysis of the pairwise kSZ power spectrum. Using the covariance matrix derived in Sec.~\ref{Sec:Covariance}, we perform a $\chi^2$ minimization to find the best fitting parameter. Since the optical depth $\tau E$ is assumed to be a proportionality constant in our template model, the $\chi^2$ minimization procedure provides the estimates for the amplitude of $\tau E$ and its associated errors as follows:
\begin{eqnarray}
	\widehat{\tau E} &=& \frac{\sum_{i,j} \widehat{P}_i C_{ij}^{-1} \bar{P}_j}{\sum_{i,j} \bar{P}_iC_{ij}^{-1} \bar{P}_j}, \nonumber \\
	\Delta (\tau E) &=& \left(\sum_{i,j} \bar{P}_iC_{ij}^{-1} \bar{P}_j \right)^{-1/2},
	\label{Eq:fitting}
\end{eqnarray}
where $C_{ij}$ is the covariance matrix estimated from the observed data itself (equations~\ref{Eq:cov_estimate} and \ref{Eq:cov_correction}), $\widehat{P}_i$ is the $i$th $k$-bin measure of $P_{ {\rm kSZ},\,\ell=1}$, and $\bar{P}_i$ is the theoretical template fixing the optical depth to $\tau E = 1$. The statistical significance of the template fit is then computed as ${\rm S/N}=|\tau E|/\Delta(\tau E)$. In our analysis, we use the full covariance matrix including the off-diagonal elements. We decide to bin the power spectrum in bins of $\Delta k = 0.02\hk$ and to use the fitting range of $0.015 \leq k \leq0.135 ~[\hk]$, thus $N_{\rm bin}=7$.

\subsection{Pairwise kSZ power spectrum dipole}

\label{Sec:ResultsPairwisePower}

We present the dipole moment of the pairwise kSZ power spectrum computed using the estimator in equation~(\ref{Eq:estimator2}). In Fig.~\ref{fig:pk}, we show our measurements of the dipole from the WPR2 CMB map with the two galaxy samples, LOWZ (red symbols) and CMASS (blue symbols), where the results at the NGC and SGC have been combined into a single measurement (equation~\ref{Eq:NGC_SGC}). 
Here, the aperture values chosen in this figure are $5\arcmin$ for LOWZ and $4\arcmin$ for CMASS, which maximize the ${\rm S/N}$ of the kSZ signal (see Table~\ref{Table:filter}).
As expected in Section~\ref{Sec:profiles}, the amplitude of the kSZ spectrum for the LOWZ sample is much larger than that for the CMASS sample, as the WPR2 map cannot resolve the CMASS halos.

We estimate the covariance error matrix of the measurement $C_{ij}$ using equations~(\ref{Eq:cov_estimate}) and (\ref{Eq:cov_correction}). Fig.~\ref{fig:pk_covariance} displays the correlation coefficient matrix, $r_{ij}=C_{ij}/\left( C_{ii}C_{jj} \right)^{1/2}$. We observe that the off-diagonal terms of the correlation coefficient are very small and their absolute values are all less than $0.25$. For LOWZ, we find the similar coefficient matrix to that for CMASS. The errorbars shown in Fig.~\ref{fig:pk} are the square root of the diagonal components, $C_{ii}^{1/2}$.

In order to see if the measured signal in Fig.~\ref{fig:pk} is due to systematic effects, we perform null tests by modifying the three-dimensional temperature fluctuation, $\delta T$ (equation~\ref{Eq:fluctuation}), in two ways for each of the two samples: we estimate $\delta T$ from (1) the WPR2 data map with the random catalogue and from (2) the WPR2 noise map with the galaxy data sample, where the galaxy density fluctuation, $\delta n$ (equation~\ref{Eq:fluctuation}), is measured from the galaxy data in all four cases. 
In case 1, we randomly select a subset of the random catalogue, so that the number of objects in the subset matches the real catalogue.
By construction these measurements should be equal to zero if there is no systematic effect. The results of these tests are shown in Fig.~\ref{fig:pk_null}. All four cases are consistent with the expectation of a null signal. We find statistical significances of ${\rm S/N}=0.32$ for LOWZ and ${\rm S/N}=0.20$ for CMASS in case $1$, and of ${\rm S/N}=0.33$ for LOWZ and $0.21$ for CMASS in case $2$.
Furthermore, we repeat the null test of case 1 three times by taking different random seeds when selecting a subset of the random catalogue.
We have found that all of these tests show ${\rm S/N}\lesssim 1$ for both LOWZ and CMASS.
We conclude from these results that there is no obvious systematic contamination affecting our analysis.

For the aperture values of $5\arcmin$ for LOWZ and $4\arcmin$ for CMASS, the optical depths $\tau E$ are respectively constrained to 
\begin{eqnarray}
	\tau E = (3.95\pm1.62)\times10^{-5} \nonumber 
\end{eqnarray}
and $\tau E = (1.25\pm1.06)\times10^{-5}$.
While for LOWZ, the statistical significance of the kSZ signal is ${\rm S/N}=2.44$,
for CMASS, the signal is consistent with null hypothesis of no signal, namely ${\rm S/N}=1.18$.
The solid lines in figure \ref{fig:pk} show the best fit fiducial models given by equation~(\ref{Eq:Template}).

\subsection{Optical depth as a function aperture radii}
\label{Sec:ResultsOpticalDepth}

\begin{table}
\centering
\begin{tabular}{cccc}
\hline\hline
AP filter $\theta_{\rm c}$ [arcmin] & $\tau E\, [10^{-5}]$ &  $\Delta(\tau E)\, [10^{-5}]$ & ${\rm S/N}$ \\
\hline
\hline
WPR2 + LOWZ \\
\hline
$\theta_{\rm c}=2$  & \hspace{-0.24cm}$-0.14$ & $1.12$ & $0.13$ \\
$\theta_{\rm c}=3$  & $1.99$  & $1.25$ & $1.59$ \\
$\theta_{\rm c}=4$  & $2.07$  & $1.32$ & $1.57$ \\
$\theta_{\rm c}=5$  & $3.95$  & $1.62$ & $2.44$ \\
$\theta_{\rm c}=6$  & $3.57$  & $1.92$ & $1.86$ \\
$\theta_{\rm c}=7$  & $4.50$  & $2.15$ & $2.09$ \\
$\theta_{\rm c}=8$  & $3.50$  & $2.40$ & $1.46$ \\
$\theta_{\rm c}=9$  & $3.01$  & $2.60$ & $1.16$ \\
$\,\,\,\theta_{\rm c}=10$ & $1.97$  & $2.73$ & $0.72$ \\
\hline
WPR2 + CMASS \\
\hline
$\theta_{\rm c}=2$  & $0.55$ & $0.94$ & $0.59$ \\
$\theta_{\rm c}=3$  & $0.95$ & $0.85$ & $1.12$ \\
$\theta_{\rm c}=4$  & $1.25$ & $1.06$ & $1.18$ \\
$\theta_{\rm c}=5$  & $1.12$ & $1.27$ & $0.88$ \\
$\theta_{\rm c}=6$  & $1.65$ & $1.63$ & $1.01$ \\
$\theta_{\rm c}=7$  & $2.02$ & $1.88$ & $1.07$ \\
$\theta_{\rm c}=8$  & $1.46$ & $2.16$ & $0.68$ \\
$\theta_{\rm c}=9$  & $2.04$ & $2.41$ & $0.85$ \\
$\,\,\,\theta_{\rm c}=10$ & $2.57$ & $2.68$ & $0.96$ \\
\hline
\end{tabular}
\caption{
Best-fit values of the measured optical depth with their associated $1\sigma$ errors for various aperture radii. The statistical significance for each aperture radius is estimated as ${\rm S/N}=|\tau E| / \Delta(\tau E)$. 
	}
\label{Table:filter}
\end{table}

To measure the optical depth $\tau E$ as a function of the angular radius of the AP filter $\theta_{\rm c}$ (Sec.~\ref{Sec:kSZ}), we repeat the analysis of the pairwise kSZ power performed in the previous subsection for various aperture radii. We summarize the resulting best-fitting values of the optical depth $\tau E$ with their $1\sigma$ errors in Table~\ref{Table:filter}. This analysis is similar to that used in~\citet{Schaan2016}, while the authors in~\citet{Schaan2016} have measured the kSZ signal by stacking the CMB temperature at the location of each halo.

Fig.~\ref{fig:tau_data_filter} shows our measurements of the optical depth $\tau E(\theta_{\rm c})$ from the WPR2 CMB map with the two galaxy samples, LOWZ (left panel) and CMASS (right panel). The best fit curves are obtained by the Gaussian gas profile model (solid lines). As expected in Sec.~\ref{Sec:kSZ}, especially for LOWZ, we find that the peak position of the AP filtered optical depth clearly shifts outward compared to that predicted by the point source assumption (dashed lines). The peak position is estimated as $\theta_{\rm peak}=2^{3/4}\sqrt{\sigma_{\rm B}^2 + \sigma_{\rm R}^2}=7.63\arcmin$ for LOWZ and $7.57\arcmin$ for CMASS (equation~\ref{Eq:Peak_position}).

The measurements of $\tau E$ for different $\theta_{\rm c}$ are strongly correlated, because the data for a smaller $\theta_{\rm c}$ is a subset of the data for a larger $\theta_{\rm c}$.
To estimate the covariance matrix of the AP filtered optical depth, we rewrite the optical depth estimator (equation~\ref{Eq:fitting}) as a function of $\theta_{\rm c}$:
\begin{eqnarray}
	\widehat{\tau E}(\theta_{\rm c}) &=& \frac{\sum_{i,j} \widehat{P}_i(\theta_{\rm c}) C_{ij}^{-1}(\theta_{\rm c}) \bar{P}_j}{\sum_{i,j} \bar{P}_iC_{ij}^{-1}(\theta_{\rm c}) \bar{P}_j},
\end{eqnarray}
where the pairwise kSZ power dipole $\widehat{P}_i(\theta_{\rm c})$ and its covariance matrix $C_{ij}(\theta_{\rm c})$ are measured from the AP filtered CMB map with the aperture radius being $\theta_{\rm c}$, and $\bar{P}$ is the theoretical template fixing the optical depth to $\tau E = 1$.
Then, the covariance of $\widehat{\tau E}(\theta_{\rm c})$ is given by
\begin{eqnarray}
	{\rm Cov}\left(  \widehat{\tau E}(\theta_a),\widehat{\tau E}(\theta_b)  \right) &=& \sum_{i,j}\sum_{k,l} {\rm Cov}\left(  \widehat{P}_i(\theta_a),\widehat{P}_k(\theta_b) \right)
	\nonumber \\
	&\times&\frac{ C_{ij}^{-1}(\theta_a) \bar{P}_j}
	{\left(\sum_{k,j} \bar{P}_iC_{ij}^{-1}(\theta_a) \bar{P}_j \right)} \nonumber \\
	&\times&
	\frac{ C_{kl}^{-1}(\theta_b) \bar{P}_l }{\left(\sum_{i,j} \bar{P}_iC_{ij}^{-1}(\theta_b) \bar{P}_j \right)},
	\label{Eq:tau_cov}
\end{eqnarray}
where 
$\theta_a\,(\theta_b)\,[{\rm arcmin}] = \{2,3,4,5,6,7,8,9,10\}$,
and ${\rm Cov}\left(  \widehat{P}_i(\theta_a),\widehat{P}_k(\theta_b) \right)$
is the covariance of $P_{ {\rm kSZ},\,\ell=1}$ between two different aperture radii, which is estimated from equation~(\ref{Eq:cov_estimate}) by replacing the variable $y_i^k$ by a function of the aperture radius $y_i^k(\theta)$.
The square root of the diagonal elements of the covariance, $\mathrm{Cov}^{ \frac{1}{2}}\left( \widehat{\tau E}(\theta),\widehat{\tau E}(\theta) \right)$ for $\theta_a=\theta_b$, reproduces the error on the optical depth at a fixed $\theta_{\rm c}$, $\Delta (\tau E)$ (equation~\ref{Eq:fitting}).

\begin{figure}
	\includegraphics[width=\columnwidth]{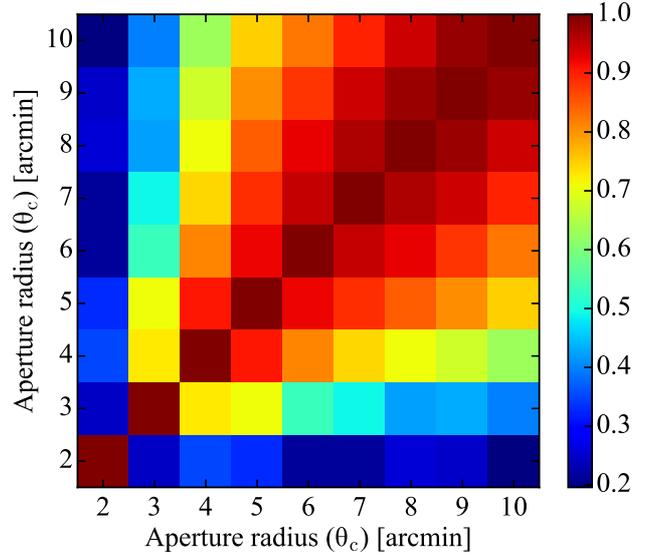}
	\caption{
Correlation coefficients of the optical depth for different aperture radii measured from the WPR2 map with the LOWZ sample. The covariance of the AP filtered optical depth presents strong correlation between different data points, especially for the largest apertures.
	}
	\label{fig:tau_covariance}
\end{figure}

Fig.~\ref{fig:tau_covariance} displays the correlation coefficient matrix for different aperture radii measured from the WPR2 map with the LOWZ sample. As expected, off-diagonal terms of the covariance present strong correlation, especially for the largest apertures. We have confirmed the similar results also for CMASS. The $1\sigma$ errorbars shown in Fig.~\ref{fig:tau_data_filter} are the square root of the diagonal elements of the covariance matrix, which are derived from equation~(\ref{Eq:fitting}). The errors increase with increasing AP filter radius $\theta_{\rm c}$, because contamination from primary CMB anisotropies increases for larger aperture radii. The grey shaded regions in Fig.~\ref{fig:tau_data_filter} are the corresponding $1\sigma$ uncertainties computed using linear theory developed in this work (see Sec.~\ref{Sec:Forecasts} and Appendix~\ref{Ap:Covariance}), which are consistent with the errorbars estimated from the data itself. 

In our analysis, we adopt the optical depth model assuming the projected Gaussian gas profile with a characteristic radius $\sigma_{\rm R}$ (equations.~\ref{tau1} and \ref{Eq:gas_profile}). We fit the measured optical depth $\tau E(\theta)$ with the template given by equation~(\ref{tau1}) with the gas fraction $f_{\rm gas}$ and the radius $\sigma_{\rm R}$ being two free parameters. The gas fraction $f_{\rm gas}$ is proportional to the kSZ signal; the radius $\sigma_{\rm R}$ characterizes how far a gas cloud can spread over a halo. In the Gaussian profile, the limit of $\sigma_{\rm R}=0$ corresponds to the assumption that observed galaxies can be treated as point sources. We do not perform analyses for other gas profiles in this work, because the errors on $\tau E$ are too large to determine the shape of detailed gas profiles, e.g. slope $\beta$ and core radius $\theta_{\rm R}$ in the $\beta$-profile given by equation~(\ref{Eq:gas_profile}).

We determine the best fitting values of $f_{\rm gas}$ and $\sigma_{\rm R}$ by minimizing
\begin{eqnarray}
	\scalebox{0.85}{$\displaystyle
	\chi^2 =  \sum_{a,b} \left[ \widehat{\tau E}(\theta_a) - \tau E^{\rm model}(\theta_a) \right]
	C_{ab}^{-1}
	\left[ \widehat{\tau E}(\theta_b) - \tau E^{\rm model}(\theta_b) \right],$}
	\label{Eq:chi2_tauE}
\end{eqnarray}
where $C_{ab}$ is the covariance matrix of $\tau E$ given by equation~(\ref{Eq:tau_cov}). We use the fitting range $\theta_{\rm c} = 2\arcmin\mathchar`-10\arcmin$ for LOWZ and $\theta_{\rm c} = 2\arcmin\mathchar`-8\arcmin$ for CMASS with $\Delta \theta_{\rm c}=1\arcmin$,
corresponding to $9$ and $7$ separation bins, respectively. We then find $(f_{\rm gas},\sigma_{\rm R})=(0.188,4.02\arcmin)$ for LOWZ and $(0.236,3.98\arcmin)$ for CMASS. 

\begin{figure}
	\includegraphics[width=\columnwidth]{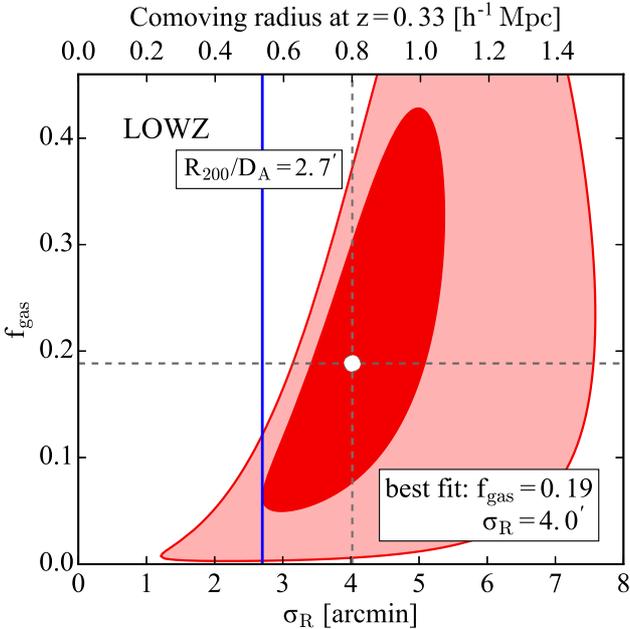}
	\caption{
Constraint on $f_{\rm gas}$ and $\sigma_{\rm R}$ for the LOWZ halos. The contours show the $\Delta \chi^2 = 2.30\, (68.3\%\, {\rm CL})$ and $\Delta \chi^2 = 6.17\, (95.4\%\, {\rm CL})$. The blue line means the virial radius $R_{\rm 200}/D_{\rm A}$. The best-fit value of the gas fraction, $f_{\rm gas}=0.188$, is consistent with the universal baryon fraction $\Omega_{\rm b}/\Omega_{\rm m}=0.155$ within the $1\sigma$ contour, and the best-fit value of the characteristic radius, $\sigma_{\rm R}=4.0\arcmin$, is consistent with the virial radius for the LOWZ halos within the $2\sigma$ contour. The point source assumption that corresponds to $\sigma_{\rm R}=0$ is ruled out at the $2.1\sigma$ level.
	}
	\label{fig:tau_contour_Gaussian}
\end{figure}

\begin{table*}
\scalebox{0.8}{
\begin{tabular}{ccccccccc}
\hline\hline
FWHM & noise & galaxy & redshift & $V$ & $\bar{n}_{\rm g}$ & $b_{\rm g}$ & $M_{\rm halo}$  & {\rm S/N}\\
$[{\rm arcmin}]$ & $[\mu K\mathchar`-{\rm arcmin}]$ &  &  & $[(h^{-1}\,{\rm Gpc})^3]$ & $[(h^{-1}\,{\rm Mpc})^{-3}]$ &  & $[h^{-1}\, M_{\sun}]$ & (${\rm FWHM}=1\arcmin,\, 1.5\arcmin,\,2.0\arcmin,\, 3.0\arcmin$)\\

\hline
\hline
CMB-S4 + BOSS\\
\hline
$1-3$ & $2$ & LRG & 0.3 $(0.15<z<0.43)$ & 2.0 & $3.0\times10^{-4}$ & 2.0 & $5.2\times10^{13}$ & $(13,\ 12,\ 12,\ 12)$ \\
$1-3$ & $2$ & LRG & 0.5 $(0.43<z<0.70)$ & 4.0 & $3.0\times10^{-4}$ & 2.0 & $2.6\times10^{13}$ & $(46,\ 43,\ 39,\ 32)$ \\
\hline
CMB-S4 + DESI\\
\hline
$1-3$ & $2$ & LRG & 0.8 $(0.65<z<0.95)$ & 13.5 & $2.8\times10^{-4}$ & 2.5 & $5.0\times10^{13}$ & $(105,\ 98,\ 90,\ 75)$ \\
$1-3$ & $2$ & ELG & 1.1 $(0.95<z<1.25)$ & 18.7 & $4.7\times10^{-4}$ & 1.4 & $4.0\times10^{12}$ & $(\ 92,\ 69,\ 52,\ 28)$ \\
\hline
CMB-S4 + PFS\\
\hline
$1-3$ & $2$ & ELG & 0.9 $(0.6<z<1.2)$ & 2.3 & $4.9\times10^{-4}$ & 1.3 & $4.0\times10^{12}$ & $(34,\ 27,\ 21,\ 12)$ \\
$1-3$ & $2$ & ELG & 1.5 $(1.2<z<2.0)$ & 4.9 & $4.7\times10^{-4}$ & 1.7 & $5.0\times10^{12}$ & $(54,\ 39,\ 30,\ 16)$ \\
\hline
\end{tabular}
}
\caption{
	Assumed experimental parameters for forecasts. 
	The first two columns show beam size and detector noise in a CMB-S4-type CMB experiment.
	Since the parameters of CMB-S4 are not yet fixed, we assume a range of beam sizes from $1\arcmin$ to $3\arcmin$.
	From the third to eighth columns show type of galaxies (LRGs or ELGs), redshift $z$, comoving survey volume $V$, 
	mean comoving number density $\bar{n}_{\rm g}$,
	linear bias parameter $b_{\rm g}$, and mean halo mass $M_{\rm halo}$.
	The last column presents the signal-to-noise ratios of the kSZ signal for some of the beam sizes (see also Fig.~\ref{fig:tau_sn}).
	The rows denote different three spectroscopic galaxy surveys, BOSS-, DESI-, and PFS-like surveys in two redshift bins.
	}
\label{Table:survey_parameters}
\end{table*}

By computing the $\chi^2$ statistics for the optical depth in equation~(\ref{Eq:chi2_tauE}) with the two varying parameters $f_{\rm gas}$ and $\sigma_{\rm R}$, we present the constraints on $f_{\rm gas}$ and $\sigma_{\rm R}$ from the WPR2 map with LOWZ, and show the result in Fig.~\ref{fig:tau_contour_Gaussian}. For CMASS, statistical uncertainties are too large to constrain on these parameters; therefore, we focus only on LOWZ. Since the amplitude of the optical depth decreases with increasing $\sigma_{\rm R}$, there is a strong degeneracy between $f_{\rm gas}$ and $\sigma_{\rm R}$. The best fit value of the gas fraction, $f_{\rm gas}=0.188$, is consistent with the universal baryon fraction $\Omega_{\rm b}/\Omega_{\rm m}=0.155$ within the $1\sigma$ contour The best fit value of the characteristic radius, $\sigma_{\rm R}=4.0\arcmin$, is slightly larger than the virial radius of the LOWZ halos, $R_{\rm 200}/D_{\rm A}=2.7\arcmin$, beyond the $1\sigma$ level but is still consistent with the virial radius within the $2\sigma$ level. We calculate the $\chi^2$ statistics in equation~(\ref{Eq:chi2_tauE}) divided by degrees of freedom (d.o.f.) with respect to the best fit value ($\chi_{\rm min}^2/{\rm d.o.f.}$); we find $\chi^2_{\rm min}/{\rm d.o.f.}=10.6/7$. We calculate $\Delta \chi^2 = \chi^2 - \chi_{\rm min}^2=6.54$ for the point source assumption ($\sigma_{\rm R}=0$). This corresponds to a probability to exceed (PTE) of $3.8\%$ or, assuming Gaussian uncertainties, a rejection of the point source assumption at the $2.1\sigma$ level.

\section{FORECASTS FOR FUTURE SURVEYS}
\label{Sec:Forecasts}

We conduct a Fisher matrix analysis (see~\citet{Tegmark1997ApJ...480...22T} for a review) to see how the next generation of high-resolution CMB experiments can improve constraints on parameters in the optical depth model (equation~\ref{tau1}). To estimate a realistic forecast for parameter constraints for a given survey, it is essential to quantify statistical errors on the observables of interest, the pairwise kSZ dipole $P_{ {\rm kSZ},\,\ell=1}(k)$ and the optical depth $\tau E(\theta)$. In this section, we present the theoretical prediction of the covariance matrix of $\tau E$, which is derived from the covariance matrix of $P_{ {\rm kSZ},\, \ell=1}$, and perform the Fisher analysis for $\tau E$ using it. We describe the full deviation of the covariance matrix of $P_{ {\rm kSZ},\, \ell=1}$ in Appendix~\ref{Ap:Covariance} and restrict the discussion here to the main results. The results we obtain depend upon a fiducial model and assumed survey parameters.

For a data vector that is predictable given parameters $p_{\alpha}$, the Fisher information matrix is:
\begin{eqnarray}
	F_{\alpha \beta} = - \left \langle \frac{\partial^2 \ln L}{\partial p_{\alpha} \partial p_{\beta}} \right\rangle,
\end{eqnarray} 
where $L$ is the likelihood function of the data set given the fiducial parameters $p_1, \cdots, p_{\alpha}$. The partial derivative with respect to parameter $p_{\alpha}$ is evaluated around the fiducial model. In the limit of Gaussian likelihood surface, the Cramer-Rao inequality shows that the Fisher matrix quantifies the best statistical errors achievable on parameter determination, marginalized over all the other parameters: $\Delta p_{\alpha} \geq (F^{-1})^{1/2}_{\alpha\alpha}$, where $(F^{-1})$ denotes the inverse of the Fisher matrix, and $\Delta p_{\alpha}$ is the relative error on $p_{\alpha}$ around its fiducial value.

\begin{figure}
	\includegraphics[width=\columnwidth]{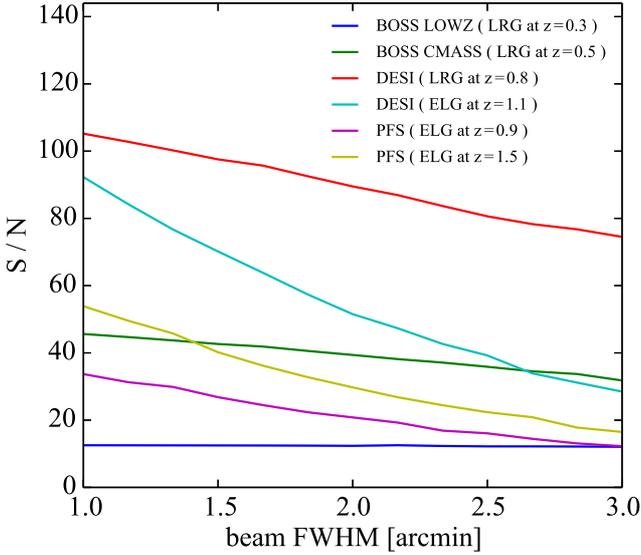}
\caption{Signal-to-noise ratios of the optical depth as a function of beam sizes for a CMB-S4-type CMB experiment with BOSS-, DESI-, and PFS-like spectroscopic galaxy surveys (For details, see Table~\ref{Table:survey_parameters}). We forecast that the DESI surveys for two types of galaxies, LRGs and ELGs, achieve ${\rm S/N}\sim100$ for ${\rm FWHM}=1\arcmin$. Since the signal of the ELGs is on scales smaller than several arc minutes, the S/N  for the ELG surveys increases with improving experimental resolution.
	}
	\label{fig:tau_sn}
\end{figure}

\begin{figure*}
	\includegraphics[width=1.0\textwidth]{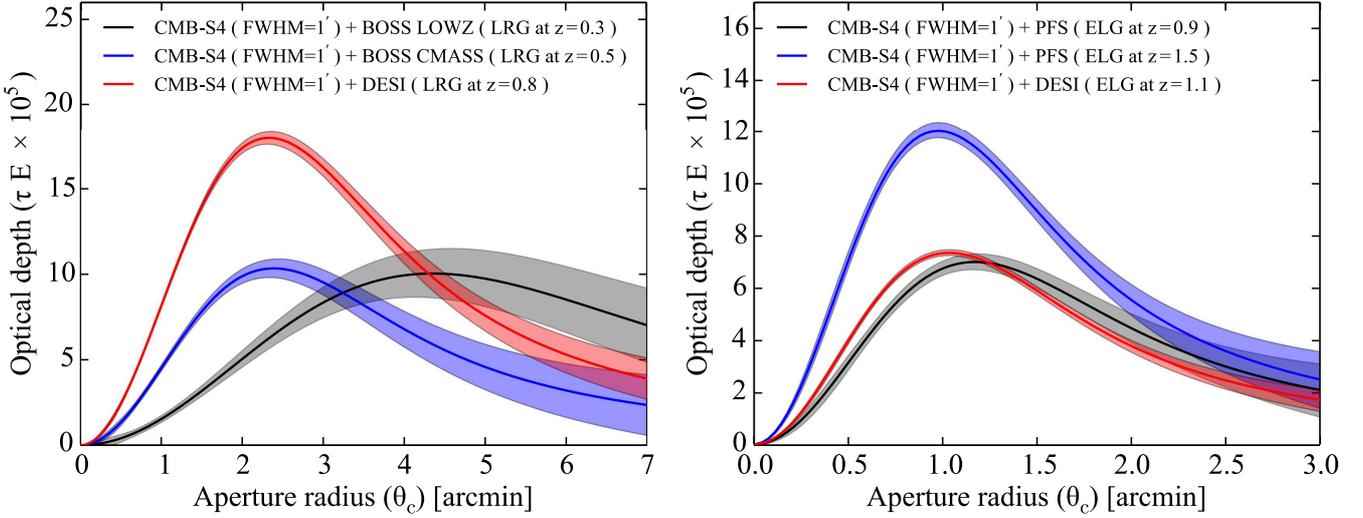}
	\caption{ 
Theoretical predictions for the optical depth as a function of the angular radius $\theta_{\rm c}$ of the AP filter, where the resolution of an assumed CMB experiment is ${\rm FWHM}=1\arcmin$. The theoretical predictions of the optical depth (solid coloured lines) are given by equation~(\ref{tau1}), which assumes a projected Gaussian gas profile, and their $1\sigma$ errors (coloured regions) are the standard deviations computed by equation~(\ref{Eq:tau_cov_theory}) in linear perturbation theory. We show the predictions of the optical depth for two types of galaxies, LRGs and ELGs, in the left panel and the right panel, respectively. A remarkable difference between LRGs and ELGs is the mass of halos hosting these galaxies: $M\sim10^{13}\,M_{\sun}$ for LRGs and $M\sim10^{12}\,M_{\sun}$ for ELGs. In the left panel, the predictions for LOWZ (black regions) and CMASS (blue regions) can be compared to the left and right panels in Fig.~\ref{fig:tau_data_filter}, respectively.
	}
	\label{fig:tau_filter}
\end{figure*}

Before computing the fisher matrix for the optical depth, we need to compute the covariance matrix of the pairwise kSZ dipole, which is given by equation~(\ref{Ap:Cov1}) in linear theory:
\begin{eqnarray}
	\scalebox{0.80}{$\displaystyle
	{\rm Cov}\left( \widehat{P}_{ {\rm kSZ},\,\ell=1}(k_1;\theta_a), \widehat{P}^*_{ {\rm kSZ},\,\ell=1}(k_2;\theta_b) \right) 
	=  \frac{\delta_{k_1k_2}^{\rm K}}{N_{\rm mode}(k_1)} C_{\rm kSZ}(k_1;\theta_a,\theta_b), $}
	\label{Eq:Cov_PkSZ}
\end{eqnarray}
where $\widehat{P}_{ {\rm kSZ},\, \ell=1}(k;\theta)$ is the pairwise kSZ power dipole estimated for a AP filter radius $\theta$, $N_{\rm mode}(k) = 4\pi k^2 \Delta k  V /(2\pi)^3$ is the number of independent Fourier modes in a bin with $\Delta k$ and $V$ being bin width and a given survey volume respectively, and $\delta^{\rm K}$ denotes the Kronecker delta such that $\delta_{k_1k_2}^{\rm K}=1$ if $k_1=k_2$ within the bin width, otherwise zero. The function $C_{\rm kSZ}(k;\theta_a,\theta_b)$ is given by equation~(\ref{Cov2}), which depends on the model of the optical depth (equation~\ref{tau1}) and the root-mean-square (rms) of the AP filtered CMB maps (equation~\ref{Ap:SigmaN})
\begin{eqnarray}
	\sigma_{\rm N}^2(\theta_a,\theta_b) = \sum_{\ell}^{\ell_{\rm max}} \frac{2\ell+1}{4\pi}
	\langle C^{\rm obs}_{\ell} \rangle U(\ell\theta_a) U(\ell\theta_b),
	\label{Eq:Sigma_N}
\end{eqnarray}
where $C_{\ell}^{\rm obs}$ is the observed CMB angular power spectrum, and $U(\ell\theta)$ is the AP filter function (equation~\ref{Eq:APfilter}). The ensemble average of $C_{\ell}^{\rm obs}$ is predictable from $\langle C^{\rm obs}_{\ell}\rangle = B_{\ell}^2C^{\rm the}_{\ell} + N_{\ell}$, where $C_{\ell}^{\rm the}$ is the theoretical prediction of the CMB power spectrum, $B_{\ell}=e^{-\sigma_{\rm B}^2\ell^2/2}$ is the Gaussian beam function, and $N_{\ell}$ denotes detector noise.

Substituting equation~(\ref{Eq:Cov_PkSZ}) into equation~(\ref{Eq:tau_cov}), we obtain the analytic form of the covariance matrix of the optical depth
\begin{eqnarray}
	{\rm Cov}\left(  \widehat{\tau E}(\theta_a),\widehat{\tau E}(\theta_b)  \right) =  \frac{1}{V}\frac{M(\theta_a,\theta_b)}{N(\theta_a)N(\theta_b)},
	\label{Eq:tau_cov_theory}
\end{eqnarray}
where
\begin{eqnarray}
	  M(\theta_a,\theta_b) &=& \int_{k_{\rm min}}^{k_{\rm max}} \frac{d^3k}{(2\pi)^3} \frac{C_{\rm kSZ}(k;\theta_a,\theta_b) [\bar{P}(k)]^2}{C_{\rm kSZ}(k;\theta_a,\theta_a)C_{\rm kSZ}(k;\theta_b,\theta_b)} \nonumber \\
	  N(\theta_a) &=& \int_{k_{\rm min}}^{k_{\rm max}} \frac{d^3k}{(2\pi)^3} \frac{[\bar{P}(k)]^2}{C_{\rm kSZ}(k;\theta_a,\theta_a)}
\end{eqnarray}
with $\bar{P}(k)$ being $P_{ {\rm kSZ},\, \ell=1}(k)$ computed for $\tau E=1$. We note that the covariance matrix of $\tau E$ does not depend on the binning of both the pairwise kSZ power spectrum and the optical depth. For the calculation of $\tau E$ for observational data, it is therefore best to choose very small bin widths, in order to minimize discretization errors. As shown in Appendix~\ref{Ap:Covariance}, the rms $\sigma_{\rm N}^2$ appears in the covariance matrix of $P_{ {\rm kSZ},\, \ell=1}$ with the snot-noise term $1/\bar{n}_{\rm g}$, where $\bar{n}_{\rm g}$ is the mean comoving number density of galaxies, so that higher values of the mean number density yield lower errors on $\tau E$. Furthermore, since the covariance scales as $1/V$ in equation~(\ref{Eq:tau_cov_theory}), larger volume surveys are suitable for the measurements of the optical depth via the kSZ effect.

Finally, we derive the Fisher matrix for the optical depth 
\begin{eqnarray}
	F_{\alpha \beta} = \sum_{a,b}\frac{\partial\, \tau E(\theta_a)}{\partial\, p_{\alpha}} 
	{\rm Cov}^{-1}\left(  \widehat{\tau E}(\theta_a),\widehat{\tau E}(\theta_b)  \right) \frac{\partial\, \tau E(\theta_b)}{\partial\, p_{\beta}},
	\label{Eq:Fisher_tau}
\end{eqnarray}
where we assumed that the covariance matrix of the optical depth is independent of the parameters $p_{\alpha}$, because we expect that the parameter dependence of the covariance becomes a sub-dominant part of the likelihood function once the parameters are sufficiently precisely determined. In what follows, we use the currently concordant flat $\Lambda$CDM model (Sec.~\ref{Sec:Introduction}). For a fair comparison with our measurements in the previous section (Fig.~\ref{fig:tau_data_filter} and~\ref{fig:tau_contour_Gaussian}), we adopt the model of the optical depth given by equation~(\ref{tau1}), which assumes a projected Gaussian gas profile with a characteristic radius estimated by the virial radius, $\sigma_{\rm R}=R_{200}/D_{\rm A}$; we assume that the gas fraction is the same as the universal baryon fraction $f_{\rm gas}=\Omega_{\rm b}/\Omega_{\rm m}=0.155$. We then allow the same two parameters to vary as those used as the fitting parameters in our analysis: $p_{\alpha} = \{f_{\rm gas},\,\sigma_{\rm R}\}$. We also calculate the expected signal-to-noise (${\rm S/N}$) ratio for $\tau E$,
\begin{eqnarray}
	\left( {\rm \frac{S}{N}} \right)^2 = \sum_{a,b} \tau E(\theta_a) 
	{\rm Cov}^{-1}(\widehat{\tau E}(\theta_a),\widehat{\tau E}(\theta_b)) \tau E(\theta_b),
	\label{Eq:tau_SN}
\end{eqnarray}
where the inverse of the ${\rm S/N}$ estimates the uncertainty of the gas fraction $f_{\rm gas}$ when all the other parameters are fixed: $\Delta f_{\rm gas}/f_{\rm gas} = ({\rm S/N})^{-1}$.

Estimating errors on the optical depth, we ignore the contribution from the pairwise kSZ octopole, as the ${\rm S/N}$ of the octopole is about ten times smaller than that of the dipole. The octopole can be used to break the degeneracy among the optical depth, the growth rate and the linear bias in the pairwise kSZ signal when we vary all these parameters~\citep{Sugiyama2016arXiv160606367S}. However, including the octopole improves the constraint on the optical depth at only a few percent level in our analysis that we fix all the other parameters.

We specify survey parameters that well resemble a CMB-S4-type CMB experiment with BOSS-, DESI-, and PFS-like spectroscopic galaxy surveys in Table~\ref{Table:survey_parameters}. Since the parameters of CMB-S4 are not yet fixed, we assume a range of beam sizes from $1\arcmin$ to $3\arcmin$
with detector noise of $2\,\mu K\mathchar`-{\rm arcmin}$. Here, we consider two types of galaxies, luminous red galaxies (LRGs) and $\left[{\rm O}\, {\rm II}\right]$ emission line galaxies (ELGs). On the one hand, LRGs lie in massive halos with $\sim 10^{13}\, h^{-1}\, M_{\sun}$ and are a potentially-useful tracer of large-scale structure for the kSZ measurement. On the other hand, ELGs are particularly useful tracers to probe baryons via the kSZ effect out to even higher redshift ($z\gtrsim1.0$) and lower halo mass ($\sim 10^{12}\, h^{-1}\, M_{\sun}$), a range of redshift and mass that is difficult to probe baryons except for the kSZ measurement. We assume that fiducial biases $b_{\rm g}$ follow constant $b_{\rm g}(z) D(z)$, where $D(z)$ is the linear growth function. For LRGs and ELGs, we use $b_{\rm LRG}D(z)=1.7$ and $b_{\rm ELG}D(z)=0.84$, respectively~\citep{Mostek_2013_2013ApJ...767...89M,DESI_Collaboration2016arXiv161100036D}. We estimate mean halo mass from the assumed bias using Press-Schechter theory~\citep{Press1974ApJ...187..425P} with Sheth-Tormen mass function~\citep{Sheth1999MNRAS.308..119S}. Computing the Fisher matrix and ${\rm S/N}$ for $\tau E$, we choose $\ell_{\rm max}=\pi/\sigma_{\rm B}$ in equation~(\ref{Eq:Sigma_N}), $k_{\rm min}=0.01\hk$ and $k_{\rm max}=0.2\hk$ in equation~(\ref{Eq:tau_cov_theory}); we adopt ranges of the AP filter of $\theta_{\rm c} = 0.5\arcmin\,\mathchar`\-\,7\arcmin$ for LRGs and $\theta_{\rm c} = 0.5\arcmin\,\mathchar`-\,3\arcmin$ for ELGs with 20 separation bins. 

We ignore contamination from the tSZ effect in equation~(\ref{Eq:Sigma_N}), because multiple frequency data in a CMB-S4 like experiment can be used to remove other foregrounds including the tSZ effect.
The tSZ cleaning may increase the noise levels significantly beyond the $2$ $\mu$K-arcmin that we use in our analysis.
We further ignore the other frequency-dependent components such as CIB, which will become dominant contribution to the expected errors on the kSZ signal for a high resolution CMB experiment.
We expect that these contaminations would make the expected ${\rm S/N}$ of the kSZ signal in this analysis smaller.
A more detail analysis is left for future works.

Fig.~\ref{fig:tau_sn} shows the ${\rm S/N}$ of the optical depth as a function of beam sizes. In the last column of Table~\ref{Table:survey_parameters}, we present the values of the ${\rm S/N}$ for some of the CMB beam sizes, ${\rm FWHM}=1\arcmin$, $1.5\arcmin$, $2.0\arcmin$, and $3.0\arcmin$. As expected, the LRG survey in DESI can achieve higher ${\rm S/N}$ than the ELG survey in DESI, as the mean halo mass for LRGs is higher than that for ELGs. However, the difference between halo masses of LRGs and ELGs, $\sim10^{13}\,M_{\sun}$ and $\sim10^{12}\,M_{\sun}$, does not become significant when CMB experiments have a sufficient resolution to resolve the host halos of a given galaxy sample, because the kSZ signal is then proportional to $M^{1/3}$ as shown in Section~\ref{Sec:profiles}. In fact, the expected ${\rm S/N}$s for LRGs and ELGs in DESI are $\sim 100$ in both types of galaxies for ${\rm FWHM}=1\arcmin$. Thus, the ${\rm S/N}$s for the ELG surveys of DESI and PFS significantly increase with decreasing beam sizes. On the other hand, the ${\rm S/N}$ for LOWZ becomes flat in a range from ${\rm FWHM}=1\arcmin$ to $3\arcmin$, because the LOWZ halos are already resolved for ${\rm FWHM}=3\arcmin$.

Fig.~\ref{fig:tau_filter} shows the theoretical predictions for the optical depth as a function of the AP filter radius $\theta_{\rm c}$, which are given by equation~(\ref{tau1}). The shaded regions are the $1\sigma$ uncertainties that are the standard deviations computed by equation~(\ref{Eq:tau_cov_theory}). To compute the predictions of the optical depth and their errors, we assume a CMB-S4 like experiment with the resolution of ${\rm FWHM}=1\arcmin$ and three spectroscopic galaxy surveys described in Table~\ref{Table:survey_parameters}. The left and right panels show the predictions for two types of galaxies, LRGs and ELGs, respectively. 

\begin{figure}
	\includegraphics[width=\columnwidth]{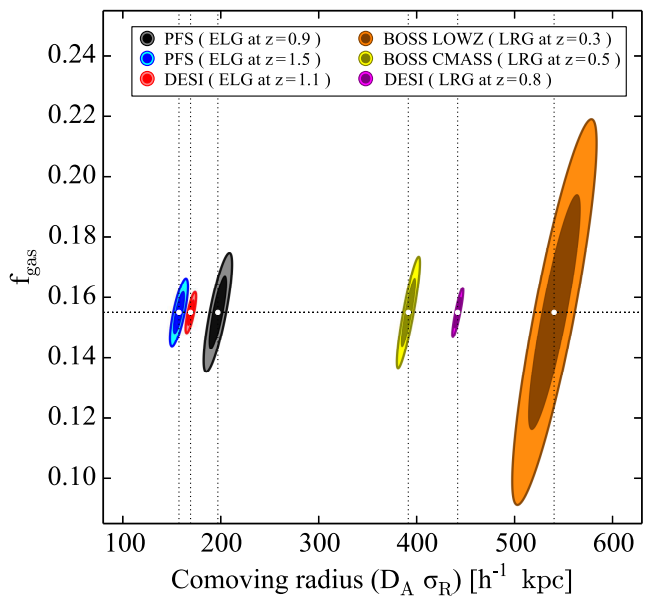}
	\caption{
Forecast confidence contours ($1$- and $2$-$\sigma$ coloured ellipses) placed on $f_{\rm gas}$ and $\sigma_{\rm R}$ derived from our Fisher matrix analysis, assuming the survey parameters of CMB-S4 with ${\rm FWHM}=1\arcmin$ and the three galaxy surveys, BOSS, DESI, and PFS, in two redshift bins. For display purposes, we plot $D_{\rm A} \sigma_{\rm R}$ on the $x$-axis, though the varying parameter is $\sigma_{\rm R}$. For DESI at $z=0.8$ (purple regions), the marginalized $1$-$\sigma$ errors on $f_{\rm gas}$ and $\sigma_{\rm R}$ are $(\Delta f_{\rm gas}/f_{\rm gas},\, \Delta \sigma_{\rm R}/\sigma_{\rm R})=(2.1\%,\,0.5\%)$. For the other surveys, see text. The prediction for LOWZ (orange regions) can be compared to Fig.~\ref{fig:tau_contour_Gaussian}.
		}
	\label{fig:tau_filter_fisher}
\end{figure}

In Fig.~\ref{fig:tau_filter_fisher}, we show the forecast constraints on $f_{\rm gas}$ and $\sigma_{\rm R}$ that are derived from the expected survey parameters of CMB-S4 with ${\rm FWHM}=1\arcmin$ and BOSS, DESI, and PFS in two redshift bins (Table~\ref{Table:survey_parameters}). In the $f_{\rm gas}-\sigma_{\rm R}$ plane, the marginalized $1\sigma$ errors on $f_{\rm gas}$ and $\sigma_{\rm R}$ are $(\Delta f_{\rm gas}/f_{\rm gas},\, \Delta \sigma_{\rm R}/\sigma_{\rm R})=(2.1\%,\,0.5\%)$ for DESI at $z=0.8$ (purple), $(4.8\%,\, 1.2\%)$ for BOSS CMASS at $z=0.5$ (yellow), $(16.6\%,\, 3.2\%)$ for BOSS LOWZ at $z=0.3$ (orange), $(1.7\%,\, 1.3\%)$ for DESI at $z=1.1$ (red), $(2.9\%,\, 2.3\%)$ for PFS at $z=1.5$ (blue), and $(5.1\%,\, 2.8\%)$ for PFS at $z=0.9$ (black).

\section{CONCLUSIONS}
\label{Sec:Conclusions}

We have presented the first measurement of the pairwise kSZ power spectrum by combining the WPR2 CMB map, which jointly uses both Planck and WMAP9 data, with the CMASS and LOWZ galaxy samples derived from BOSS DR12. We apply an aperture photometry filter to estimate the CMB temperature distortion associated with each galaxy and measure the filtered optical depth of the host halos of a given galaxy sample via the pairwise kSZ signal. By repeating the analyses of the pairwise kSZ power in the filtered CMB maps for various aperture radii, we measure the optical depth as a function of the aperture radius and constrain the gas fraction, $f_{\rm gas}$, as well as a parameter of an assumed Gaussian gas profile, the characteristic radius $\sigma_{\rm R}$. This analysis is a direct measurement of baryonic matter in the Universe that is insensitive to the gas temperature. While the ${\rm S/N}$ of the current measurement is low, this paper both illustrates the methodology for future surveys and places constraints on the optical depth.

Table~\ref{Table:filter} summarizes our constraints on the optical depth for various aperture radii, and Fig.~\ref{fig:tau_data_filter} plots them. In our analysis, we constrain not the optical depth $\tau$ itself but the product $\tau E$, where $E=H/H_0$ is the ratio between the Hubble parameters at redshift $z$ and the present day. For LOWZ, we find $\tau E = (3.95\pm1.62)\times10^{-5}$ for $5\arcmin$ aperture, corresponding to a statistical significance of ${\rm S/N}=2.44$. For CMASS, the kSZ signal is consistent with null hypothesis of no signal. Fig.~\ref{fig:tau_contour_Gaussian} shows the degeneracy between the gas fraction and the characteristic radius of the Gaussian gas profile with their best fitting values, $(f_{\rm gas},\, \sigma_{\rm R})=(0.19,\, 4.0\arcmin)$, for LOWZ. This result implies that the gas fraction is consistent with the universal baryon fraction $f_{\rm b}=\Omega_{\rm b}/\Omega_{\rm m}=0.155$ within the $1\sigma$ level, and that the gas radius, which is roughly estimated by $\sigma_{\rm R} D_{\rm A}$ with the angular diameter distance $D_{\rm A}$, is consistent with the virial radius $R_{\rm 200}$ within the $2\sigma$ level. The point source assumption in the model of the gas profile is ruled out at the $2.1\sigma$ level for the LOWZ halos, implying the presence of the spread of the free electron gas in the LOWZ halos. While we have analyzed the projected Gaussian gas profile in this paper, we defer a more detailed study of gas profile models for future kSZ works~(e.g., \cite{Flender2016arXiv161008029F}).

To illustrate the constraining power of the kSZ measurement on the optical depth, we provide a forecast of future kSZ measurements by performing the Fisher analysis. For the computation of the Fisher matrix, we assume a CMB-S4-type CMB experiment with BOSS-, DESI-, and PFS-like spectroscopic galaxy surveys. These galaxy surveys measure two types of galaxies, LRGs and ELGs, which lies in halos with masses of $\sim10^{13}\, M_{\sun}$ and $\sim10^{12}\, M_{\sun}$ and at redshifts of $z\lesssim1.0$ and $z\gtrsim1.0$, respectively. Thus, the future measurements of the kSZ signal for various galaxy surveys will allow one to constrain the properties of the free electron gas within halos for a broad range of mass and redshift. In particular, both the LRG and ELG surveys in DESI will achieve detections of the kSZ signal with ${\rm S/N}\sim100$ for CMB-S4 assuming ${\rm FWHM}=1\arcmin$. This forecast strengthens the motivation for CMB-S4 making small scale measurement of CMB temperature fluctuations over regions of the sky covered by upcoming redshift surveys. While this paper emphasized the upcoming the CMB-S4 experiment and the DESI and PFS surveys, we anticipate that we would reach similar conclusions for the CMB experiment of Advanced ACTPol~\citep{Calabrese:2014gwa,Henderson:2015nzj} and the spectroscopic surveys of the ESA satellite mission Euclid~\citep{Laureijs2011arXiv1110.3193L,Amendola2016arXiv160600180A} and NASA's Wide Field Infrared Survey Telescope (WFIRST;~\cite{Spergel2013arXiv1305.5425S}).

\section*{ACKNOWLEDGEMENTS}

We are grateful to Chiaki Hikage, Maresuke Shiraishi, Akito Kusaka, Shun Saito, Hironao Miyatake, Masamune Oguri, Daisuke Nagai, Masato Shirasaki, and Eiichiro Komatsu for very useful discussion. NSS acknowledges financial support from Grant-in-Aid for JSPS Fellows (No. 28-1890). NSS further acknowledges financial support from Grant-in-Aid for Scientific Research from the JSPS Promotion of Science (25287050) and from a NEXT project ``Priority issue 9 to be tackled by using post-K computer''.  The Flatiron Institute is supported by the Simons Foundation.  Numerical computations were carried out on Cray XC30 at Center for Computational Astrophysics, National Astronomical Observatory of Japan. We thank the LGMCA team for publicly releasing their CMB maps. Some of the results in this paper have been derived using the \textit{HEALPix} package~\citep{Gorski2005ApJ...622..759G}. Linear matter power spectra and CMB temperature power spectra used in this paper are generated with \textit{CLASS}~\citep{Lesgourgues:2011re}.

Funding for SDSS-III has been provided by the Alfred P. Sloan Foundation, the Participating Institutions, the National Science Foundation, and the U.S. Department of Energy Office of Science. The SDSS-III web site is \url{ttp://www.sdss3.org/}. SDSS-III is managed by the Astrophysical Research Consortium for the Participating Institutions of the SDSS-III Collaboration including the University of Arizona, the Brazilian Participation Group, Brookhaven National Laboratory, Carnegie Mellon University, University of Florida, the French Participation Group, the German Participation Group, Harvard University, the Instituto de Astrofisica de Canarias, the Michigan State/Notre Dame/JINA Participation Group, Johns Hopkins University, Lawrence Berkeley National Laboratory, Max Planck Institute for Astrophysics, Max Planck Institute for Extraterrestrial Physics, New Mexico State University, New York University, Ohio State University, Pennsylvania State University, University of Portsmouth, Princeton University, the Spanish Participation Group, University of Tokyo, University of Utah, Vanderbilt University, University of Virginia, University of Washington, and Yale University. 

Based on observations obtained with Planck (\url{http://www.esa.int/Planck}), an ESA science mission with instruments and contributions directly funded by ESA Member States, NASA, and Canada.

\bibliographystyle{mnras}
\bibliography{ms} 

\appendix

\section{COVARIANCE MATRICES}
\label{Ap:Covariance}

While there are a few analytic approaches for the covariance of pairwise velocities~\citep{Bhattacharya2008,Mueller2015a}, they do not include contamination of the kSZ signal, e.g. primary CMB anisotropies and detector noise. In~\citet{Sugiyama2016arXiv160606367S}, we have presented the model of the covariance of the pairwise kSZ power multipoles, which may include the contamination. In this appendix, to validate our covariance model, we compare our model with covariance estimates from data itself using the jackknife (JK)~\citep{Quenouille1956,Tukey1958} and bootstrap (BS)~\citep{Efron1979} methods. Section~\ref{Ap:Theory} presents the model of the covariance in linear theory, and Section~\ref{Ap:Comparison} compares the model with the measurement from the data.

\subsection{Theory}
\label{Ap:Theory}

We estimate the CMB temperature distortion associated with each galaxy by applying the AP filter (Sec.~\ref{Sec:kSZ}). Then, we model the observed CMB temperature in the AP filtered CMB map as follows
\begin{eqnarray}
	\delta T_{\rm AP}(\hat{n}_i;\theta_{\rm c}) = 
	\delta T_{\rm kSZ}^{(\rm AP)}(\hat{n}_i;\theta_{\rm c}) + \delta T^{(\rm AP)}_{ {\rm N}}(\hat{n}_i;\theta_{\rm c}), 
\end{eqnarray}
where $\theta_{\rm c}$ denotes the AP filter radius, the line-of-sight $\hat{n}_i$ points to galaxy $i$ from the observer,
$\delta T^{\rm (AP)}_{\rm kSZ} = -(T_0\tau/c)\, \vec{v}\cdot\hat{n}$ (equation~\ref{Eq:kSZ_tau}) represents the kSZ signal, and $\delta T^{\rm (AP)}_{\rm N}$ is the total noise on the kSZ signal, which may include primary CMB anisotropies, detector noise, thermal SZ effect, and residual foregrounds. Since we measure the CMB temperature distortion subtracting its average (equation~\ref{Eq:DeltaT_AP}), the ensemble averages of $\delta T_{\rm kSZ}^{\rm (AP)}$ and $\delta T_{\rm N}^{\rm (AP)}$ will be zero: $\langle \delta T_{\rm kSZ}^{\rm (AP)}\rangle=\langle \delta T^{\rm (AP)}_{\rm N}\rangle = 0$.

The pairwise kSZ power spectrum (equation~\ref{Eq:PairwiseKSZPower}) including the noise $\delta T^{\rm (AP)}_{\rm N}$ for aperture radius $\theta_{\rm c}$ is given by
\begin{eqnarray}
	&& 
	\scalebox{0.72}{$\displaystyle
	\hspace{-2.0cm}P_{\rm kSZ}(\vec{k};\theta_{\rm c}) 
	=\left\langle-\frac{V}{N^2}\sum_{i,j} \left[  \delta T_{\rm kSZ}^{(\rm AP)}(\hat{n}_i;\theta_{\rm c}) - \delta T_{\rm kSZ}^{(\rm AP)}(\hat{n}_j;\theta_{\rm c})  \right] 
	e^{-i\vec{k}\cdot(\vec{s}_i-\vec{s}_j)} \right\rangle $}
	\nonumber \\
	&&
	\scalebox{0.72}{$\displaystyle
	\hspace{-0.3cm}
	+\left\langle-\frac{V}{N^2}\sum_{i,j} \left[  \delta T_{\rm N}^{(\rm AP)}(\hat{n}_i;\theta_{\rm c}) - \delta T_{\rm N}^{(\rm AP)}(\hat{n}_j;\theta_{\rm c})  \right] 
	e^{-i\vec{k}\cdot(\vec{s}_i-\vec{s}_j)} \right\rangle,
	$}
	\label{Ap:pk}
\end{eqnarray}
where $\vec{s}_i$ denotes the observed coordinates of galaxy $i$, $N$ represents the total number of observed galaxies, and $V$ is survey volume. In the pairwise kSZ signal, the contribution from the noise in the second line of equation~(\ref{Ap:pk}) becomes zero, because its parity is even, while the parity of the kSZ signal is odd (Sec.~\ref{Sec:Estimator}). 

The kSZ power covariance can be formally expressed in terms of Gaussian (unconnected) and non-Gaussian (connected) contributions in cumulant expansion. The Gaussian term has only diagonal elements of the covariance matrix, and the power spectrum estimates of different scales are uncorrelated, while non-vanishing off-diagonal elements arise from the non-Gaussian term on small scales. Linear theory only yields the Gaussian term, described as
\begin{eqnarray}
	\scalebox{0.80}{$\displaystyle
	{\rm Cov}\left( \widehat{P}_{ {\rm kSZ},\,\ell=1}(k_1;\theta_a), \widehat{P}^*_{ {\rm kSZ},\,\ell=1}(k_2;\theta_b) \right) 
	=  \frac{\delta_{k_1k_2}^{\rm K}}{N_{\rm mode}(k_1)} C_{\rm kSZ}(k_1;\theta_a,\theta_b), $}
	\label{Ap:Cov1}
\end{eqnarray}
where $\widehat{P}_{ {\rm kSZ},\, \ell=1}(k;\theta)$ is the pairwise kSZ power dipole estimated for aperture radius $\theta$ (equation~\ref{Eq:estimator1}), $N_{\rm mode}=4\pi k^2\Delta kV/(2\pi)^3$ is the number of independent Fourier modes in a bin, $\Delta k$ is the bin width, and $\delta^{ {\rm K}}$ denotes the Kronecker delta such that $\delta_{k_1k_2}^{\rm K}=1$ if $k_1=k_2$ within the bin width, otherwise zero.

\begin{figure}
	\includegraphics[width=\columnwidth]{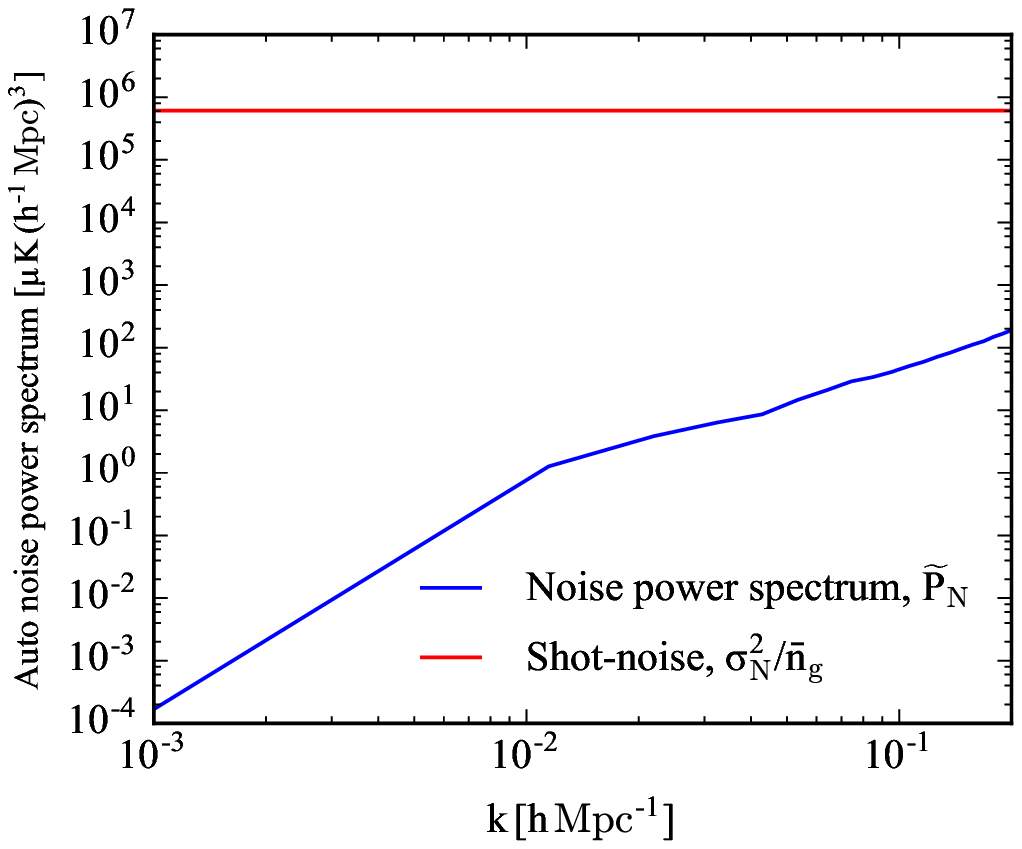}
	\caption{
Auto power spectrum of the noise fluctuation $\widetilde{P}_{\rm N}$ (blue line) and its shot-noise term $\sigma_{\rm N}^2/\bar{n}_{\rm g}$ (red line) in the AP filtered WPR2 map (Sec.~\ref{Sec:Data_CMB}). The AP filtered $\widetilde{P}_{\rm N}$ is about $10^{3}$ times smaller than the shot-noise $\sigma_{\rm N}^2/\bar{n}_{\rm g}$ at the scales of interest $(k<0.2\hk)$; therefore, we can make an approximation that the AP filtered noise fluctuation $\delta T_{\rm N}^{(\rm AP)}$ is an uncorrelated, Gaussian field (equation~\ref{Eq:noise}) on large scales.
	}
	\label{fig:NoisePowerSpectrum}
\end{figure}

\begin{figure*}
	\includegraphics[width=1.0\textwidth]{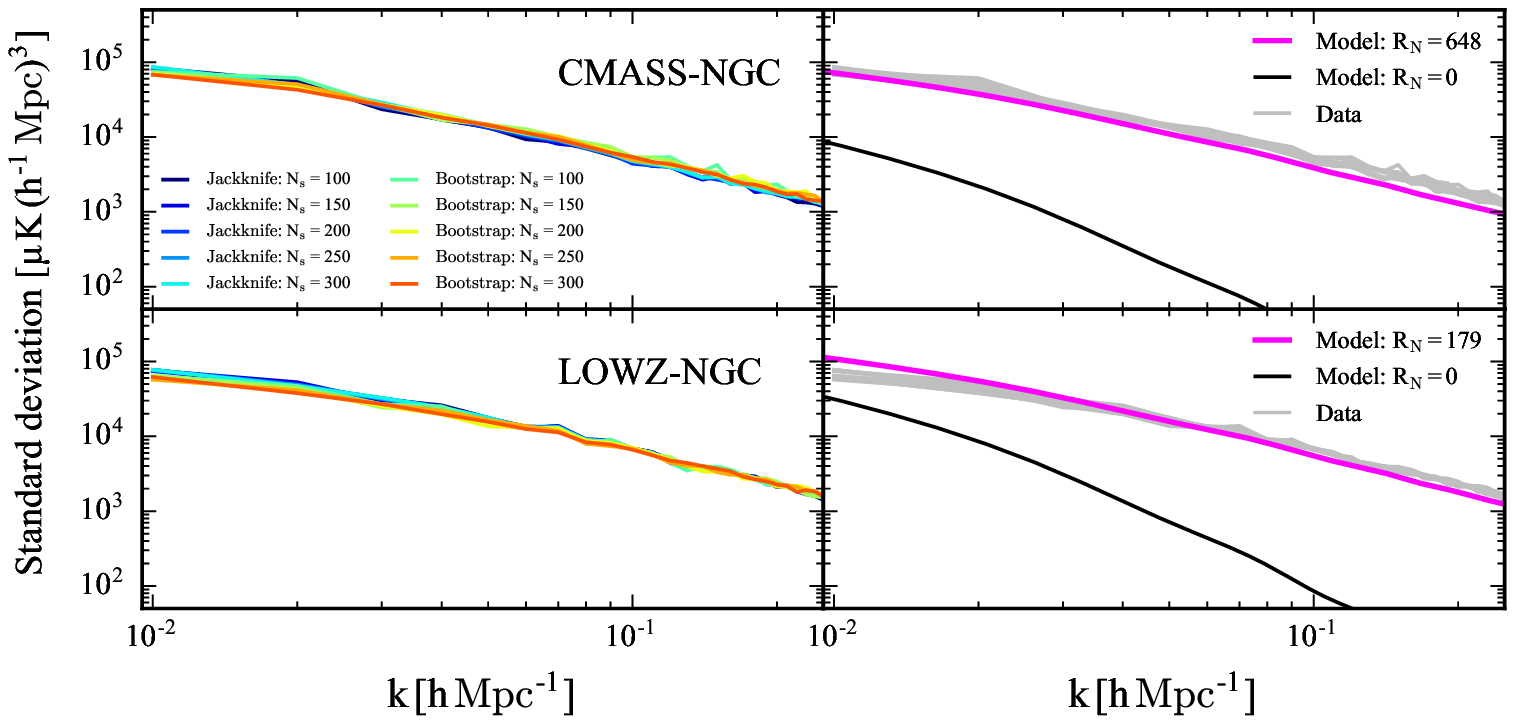}
	\caption{
Left: Standard deviations of the pairwise kSZ power dipole as a function of scales for different internal error estimates using the jackknife and bootstrap methods for various spatial sub-regions on the sky ($N_{\rm s}=100,\, 150,\, 200,\, 250,\, {\rm and}\, 300$). Data used here are the WPR2 map for $5\arcmin$ aperture with the LOWZ-NGC (left bottom panel) and CMASS-NGC (left top panel) samples. Right: Comparisons of the internal error estimates with the predictions from our model in equation~(\ref{Cov2}). The grey lines are the same as the coloured lines in the left panel. The black and magenta solid lines show the linear theory predictions for a noise less sky map ($R_{\rm N}=0$) and the WPR2 map ($R_{\rm N}=179$ for LOWZ and $R_{\rm N}=648$ for CMASS), respectively. The values of the inverse signal-to-noise ratio $R_{\rm N}$ are estimated using the WPR2 map and the linear theory predictions of the LOS velocity dispersion (see text for further details). Our model is consistent with the internal error estimates, which means that the AP filtered noise properties on the kSZ signal can be explained by an uncorrelated, Gaussian field (equation~\ref{Eq:noise}).
	}
	\label{fig:StandardDeviation}
\end{figure*}

Since the AP filtering removes most of any large-scale noise contributions, we assume that the AP filtered noise fluctuation $\delta T_{\rm N}^{\rm (AP)}$ is an uncorrelated, Gaussian field that satisfies
\begin{eqnarray}
	\langle \delta T_{ {\rm N}}^{(\rm AP)}(\hat{n}_i;\theta_a) \delta T^{(\rm AP)}_{ {\rm N}}(\hat{n}_j;\theta_b) \rangle_{\rm c} 
	= \sigma_{\rm N}^2(\theta_a,\theta_b) \delta_{ij}, 
	\label{Eq:noise}
\end{eqnarray}
where $\langle \cdots \rangle_{\rm c}$ denotes a cumulant. In the above expression, the root-mean-square (rms) of the noise fluctuation, $\sigma_{\rm N}$, is given by
\begin{eqnarray}
	\sigma_{\rm N}^2(\theta_a,\theta_b) = \sum_{\ell} \frac{2\ell+1}{4\pi} C_{\ell} U(\ell \theta_a)U(\ell \theta_b),
	\label{Ap:SigmaN}
\end{eqnarray}
where $C_{\ell}$ is the CMB angular power spectrum, and $U_{\ell}$ is the AP filter function (equation~\ref{Eq:APfilter}). On the other hand, the rms of the kSZ signal, which represents a typical kSZ temperature, is given by
\begin{eqnarray}
	\sigma_{\rm kSZ}(\theta_{\rm c}) = \left( \frac{T_0\tau(\theta_{\rm c})}{c} \right)\sigma_{\rm v},
\end{eqnarray}
where the optical depth $\tau (\theta_{\rm c})$ as a function of aperture radii is given by equation~(\ref{tau1}), and $\sigma_{\rm v}$ is the LOS velocity dispersion, which is computed by $\sigma_{\rm v} = \left( aHf\sigma_8 \right) \sigma_{\rm 0}$ with $\sigma_{0}^2 =  \frac{1}{3}\int \frac{dk}{2\pi^2} P_{\rm lin}(k)$ in linear theory. Then, to estimate the impact of the noise level, we define the following inverse signal-to-noise ratio as
\begin{eqnarray}
	R^2_{\rm N}(\theta_a,\theta_b) = \frac{\sigma^2_{\rm N}(\theta_a,\theta_b)}{\sigma_{\rm kSZ}(\theta_a)\sigma_{\rm kSZ}(\theta_b)}.
\end{eqnarray}
Under the assumption in equation~(\ref{Eq:noise}), we finally calculate $C_{\rm kSZ}$ in linear theory (see equation~(4.9) in~\citet{Sugiyama2016arXiv160606367S}):
\begin{eqnarray}
	&&
	\scalebox{0.85}{$\displaystyle
 	\hspace{-0.8cm}C_{ {\rm kSZ}}(k;\theta_a,\theta_b) 
	$}
	 \nonumber \\
	&&
	\scalebox{0.85}{$\displaystyle
	\hspace{-0.8cm}= 18 \left( \frac{T_0H_0}{c} \right)^2\, a^2E^2\,\tau(\theta_a) \tau(\theta_b)\,  (f\sigma_8)^2$}
	\nonumber \\
	&& 
	\scalebox{0.85}{$\displaystyle
	\hspace{-0.8cm}\Bigg[ \left( \frac{4}{5}(b\sigma_8)^2  + \frac{8}{7}(b\sigma_8)(f\sigma_8) 
		  + \frac{4}{9}(f\sigma_8)^2 \right) \frac{(P_{\rm lin}(k))^2}{k^2}
	$}
	\nonumber \\
	 &&
	\scalebox{0.85}{$\displaystyle
	\hspace{-0.6cm} +  \frac{\sigma_0^2}{\bar{n}_{\rm g}}\left( 1 + R_{\rm N}^2(\theta_a,\theta_b)  \right)
		  \left( \frac{2}{3}(b\sigma_8)^2  + \frac{4}{5}(b\sigma_8)(f\sigma_8) + \frac{2}{7}(f\sigma_8)^2 \right) P_{\rm lin}(k)
	$}
	\nonumber \\
	&&
	\scalebox{0.85}{$\displaystyle
	\hspace{-0.6cm} + \frac{2}{5} \frac{1}{\bar{n}_{\rm g}} \frac{P_{\rm lin}(k)}{k^2}
	+ \frac{2}{3}\frac{\sigma_0^2}{\bar{n}_{\rm g}^2}\left( 1 + R_{\rm N}^2(\theta_a,\theta_b) \right)
	\Bigg],
	$}
	  \label{Cov2}
\end{eqnarray}
where $\bar{n}_{\rm g}$ is the galaxy mean number density. The above expression holds true even if $\langle \delta T_{\rm N}^{\rm (AP)}\rangle\neq 0 $ as shown in~\citet{Sugiyama2016arXiv160606367S}. 

To validate the assumption of the noise property in equation~(\ref{Eq:noise}), we calculate the auto power spectrum of the noise fluctuation $\delta T_{\rm N}$, which mainly contributes to the covariance of the pairwise kSZ dipole, without the assumption. In linear theory, we have
\begin{eqnarray}
	\scalebox{0.85}{$\displaystyle
	P_{\rm N}(\vec{k};\theta_a,\theta_b) $}
	&\scalebox{0.85}{$\displaystyle = $}& 
	\scalebox{0.85}{$\displaystyle
	  \frac{V}{N^2}\sum_{i,j} 
	\left\langle \delta T_{\rm N}^{(\rm AP)}(\hat{n}_i;\theta_a) T_{\rm N}^{(\rm AP)}(\hat{n}_j;\theta_b) \right\rangle_{\rm c}
	e^{-i\vec{k}\cdot(\vec{s}_i-\vec{s}_j)},$}
	\nonumber \\
	&\scalebox{0.85}{$\displaystyle = $}& 
	\scalebox{0.85}{$\displaystyle
   	\widetilde{P}_{\rm N}(\vec{k};\theta_a,\theta_b) + \frac{\sigma_{\rm N}^2(\theta_a,\theta_b)}{\bar{n}_{\rm g}} $},
\end{eqnarray}
where $\widetilde{P}_{\rm N}$ is given by
\begin{eqnarray}
	\widetilde{P}_{\rm N}(\vec{k};\theta_a,\theta_b) &=&  \frac{1}{V} \sum_{\ell m} C_{\ell} U(\ell \theta_a)U(\ell \theta_b) Y_{\ell}^m (\hat{k})Y_{\ell}^{m*}(\hat{k}) \nonumber \\
	&& \left[  4\pi\int_0^{x_{\rm max}} dx x^2  j_{\ell}(kx)  \right]^2,
\end{eqnarray}
and the shot-noise term $\sigma_{\rm N}^2/\bar{n}_{\rm g}$ appears in our covariance model in equation~(\ref{Cov2}). Since the assumption in equation~(\ref{Eq:noise}) only yields the shot-noise term, we show that $\widetilde{P}_{\rm N}$ is small enough to be ignored compared to $\sigma_{\rm N}^2/\bar{n}_{\rm g}$. Fig.~\ref{fig:NoisePowerSpectrum} shows $\widetilde{P}_{\rm N}$ (blue line) and $\sigma_{\rm N}^2/\bar{n}_{\rm g}$ (red line) for $\theta_a=\theta_b=5\arcmin$, $\bar{n}_{\rm g} = 3.0\times 10^{-4}\, h^{3}\, {\rm Mpc}^{-3}$, and $V = (4\pi/3)x_{\rm max}^3 = 1\, h^{-1}\, {\rm Gpc}$. The CMB angular power spectrum $C_{\ell}$ is computed from the WPR2 map we use in this work (Sec.~\ref{Sec:Data_CMB}), which includes all of the noise components. As expected, we find that the shot-noise term $\sigma_{\rm N}^2/\bar{n}_{\rm g}$ is dominant, and the AP filtered noise power spectrum $\widetilde{P}_{\rm N}$ is about $10^{3}$ times smaller than the shot noise at the scales of interest ($k<0.2\hk$).

\subsection{Comparisons with data}
\label{Ap:Comparison}

We estimate the data covariance matrices through the resampling of data itself, using jackknife~\citep{Quenouille1956,Tukey1958} and bootstrap~\citep{Efron1979} methods. The galaxy dataset has been split into $N_{\rm s}$ spatial sub-regions on the sky. For the jackknife method, a copy of data is defined by systematically omitting, in turn, each of $N_{\rm s}$ sub-regions. For the bootstrap method, $N_{\rm s}$ sub-regions are selected at random with replacement from the initial data set. 

In the jackknife and bootstrap schemes, the covariance matrices are estimated by
\begin{eqnarray}
	C_{ {\rm JK}, ij} &=& \frac{N_{\rm s}-1}{N_{\rm s}} 
	  \sum_{k}^{N_{\rm s}} \left( y^k_i - \bar{y}_i \right) \left( y^k_j - \bar{y}_j \right) \nonumber \\
	  C_{ {\rm BS}, ij} &=& \frac{1}{N_{\rm s}-1} 
	  \sum_{k}^{N_{\rm s}} \left( y^k_i - \bar{y}_i \right) \left( y^k_j - \bar{y}_j \right),
\end{eqnarray}
where $y_i^k$ is the $i$th measure of the statistic of interest for the $k$th configuration, and $\bar{y}_i = \frac{1}{N_{\rm s}} \sum_k^{N_{\rm s}} y_i^k$. Here, we choose $y_i=P_{\rm kSZ,\ell=1}(k_i)$.

We calculate the rms noise $\sigma_{\rm N}(\theta_{\rm c})=\sqrt{\sigma_{\rm N}^2(\theta_{\rm c},\theta_{\rm c})}$ in equation~(\ref{Ap:SigmaN}) from the WPR2 map for $5\arcmin$ aperture; we find $\sigma_{\rm N}=13.6\,\mu K$. In linear theory, the LOS velocity dispersion is $\sigma_{\rm v}=253\kms$ for LOWZ and $\sigma_{\rm v}=281\kms$ for CMASS, which correspond to typical kSZ temperatures of $\sigma_{\rm kSZ}=0.076\,\mu K$ and $\sigma_{\rm kSZ}=0.021\,\mu K$, respectively, where we used the best fit values of the optical depth $\tau E$ for $5\arcmin$ aperture in Table~\ref{Table:filter}. As a result, we evaluate inverse signal-to-noise ratios of $R_{\rm N} = 179$ for LOWZ  and $R_{\rm N}=648$ for CMASS.

The left panels of Fig.~\ref{fig:StandardDeviation} show the standard deviations of the pairwise kSZ power dipole for different internal error estimates using the jackknife and bootstrap methods for various spatial sub-regions on the sky (coloured lines). We use the WPR2 map for $5\arcmin$ aperture with the LOWZ-NGC (left bottom panel) and CMASS-NGC (left top panel) samples. In the right panels, we compare the internal error estimates (grey lines) with the predictions from our model in equation~(\ref{Cov2}), where we use $\bar{n}_{\rm g} = 3.6\, (2.6)\times 10^{-4}\, (\hMpc)^{-3}$, $V = 0.90\, (2.57)\, (h^{-1}\, {\rm Gpc})^3$, and $R_{\rm N}=179\, (648)$ for the LOWZ-NGC (CMASS-NGC) sample. The grey lines in the right panels are the same as the coloured lines in the left panels, and the black and magenta solid lines are the linear theory predictions for a noise less sky map ($R_{\rm N}=0$) and the WPR2 map ($R_{\rm N}=179$ for LOWZ and $R_{\rm N}=648$ for CMASS), respectively. We can find a excellent agreement between the covariance estimates from the model and the data, indicating that the property of the AP filtered noise on the kSZ signal can be explained as an uncorrelated, Gaussian field (equation~\ref{Eq:noise}).

\section{Details of how to measure the pairwise kSZ power spectrum}
\label{Ap:Details}

In this appendix, we detial how to measure the pairwise kSZ power spectrum from galaxy catalogues.

\subsection{Estimator}
\label{Ap:Estimator}

We start with rewriting equation~(\ref{Eq:estimator1}):
\begin{eqnarray}
	  \widehat{P}_{\rm kSZ, \ell}(\vec{k})
	&=&  - \frac{\left( 2\ell+1 \right)}{A}
	\int d^3s_1 \int d^3s_2\, e^{-i\vec{k}\cdot\vec{s}_{12}}
	{\cal L}_{\ell}(\hat{k}\cdot\hat{n}_{12}) \nonumber \\
	&\times& \Big[ \delta T(\vec{s}_1) \delta n(\vec{s}_2) - \delta n(\vec{s}_1)\delta T(\vec{s}_2) \Big].
\end{eqnarray}
As the integrals over $\vec{s}_1$ and $\vec{s}_2$ in the above expression are not separable due to the $\hat{n}_{12}$ dependence in ${\cal L}_{\ell \geq 1}( \hat{k}\cdot\hat{n}_{12} )$, the computation of the estimator will be ${\cal O}(N_{k} \times N^2)$, where $N_k$ and $N$ are the numbers of $k$-modes and observed galaxies, respectively. Since the estimator is computationally challenging, we provide a fast method to measure the pairwise kSZ power multipoles using the local plane parallel approximation,
\begin{eqnarray}
	{\cal L}_{\ell}(\hat{k}\cdot\hat{n}_{12}) \approx {\cal L}_{\ell}(\hat{k}\cdot\hat{s}_1) \approx {\cal L}_{\ell}(\hat{k}\cdot\hat{s}_2).
\end{eqnarray}
This approximation allows the integrals in equation~(\ref{Eq:estimator1}) to decouple into a product of Fourier transforms~\citep{Yamamoto2006PASJ...58...93Y,Blake2011MNRAS.415.2876B,Beutler2014MNRAS.443.1065B,Samushia2015MNRAS.452.3704S}, which reduces the process to ${\cal O}(N_k \times N)$:
\begin{eqnarray}
	  \widehat{P}_{\rm kSZ, \ell}(\vec{k})
	=  - \frac{\left( 2\ell+1 \right)}{A} \left[ \delta T_{\ell}(\vec{k})\, \delta n_0^*(\vec{k})  -\mathrm{c.c.}\right],
\end{eqnarray}
where $\delta n(\VEC{k})$ is the Fourier transform of the density fluctuation $\delta n(\vec{s})$:
\begin{eqnarray}
	\delta n(\vec{k}) = \int d^3s\, e^{-i\vec{k}\cdot\vec{s}} \delta n(\vec{s}\,),
\end{eqnarray}
and $\delta T_{\ell}(\vec{k})$ is the multipole component of the kSZ temperature fluctuation $\delta T(\vec{s})$, given by
\begin{eqnarray}
	  \delta T_{\ell}(\vec{k}) = \int d^3s\, e^{-i\vec{k}\cdot\vec{s}} {\cal L}_{\ell}(\hat{k}\cdot\hat{s}) \delta T(\vec{s}\,).
\end{eqnarray}
This approximation has been known to be accurate on the scales of interest for the galaxy power spectrum ($k\lesssim 0.2\hk$)~\citep{Samushia2015MNRAS.452.3704S,Yoo2015MNRAS.447.1789Y}, and we expect that it can be also applied to the pairwise kSZ power spectrum. Furthermore, as pointed out in \citep{Bianchi2015MNRAS.453L..11B,Scoccimarro2015PhRvD..92h3532S}, $\delta T_{\ell}(\vec{k})$ can be built to be computable using any fast Fourier transform (FFT) algorithm, leading to ${\cal O}(N_k \log N_k)$: e.g. for $\ell=1$ and $\ell=3$, we have
\begin{eqnarray}
	  \delta T_{\ell=1}(\vec{k}) &=& \hat{k}_i \delta T_{i}(\vec{k}), \nonumber \\
	  \delta T_{\ell=3}(\vec{k}) &=& \frac{5}{2}\hat{k}_i \hat{k}_j \hat{k}_k \delta T_{ijk}(\vec{k}) - \frac{3}{2}\hat{k}_i \delta T_{i}(\vec{k}),
\end{eqnarray}
with
\begin{eqnarray}
	  \delta T_{i_1 \dots i_n}(\vec{k}) &=&  \int d^3s\, e^{-i\vec{k}\cdot\vec{s}} \hat{s}_{i_1}\cdots\hat{s}_{i_n} \delta T(\vec{s}\,),
\end{eqnarray}
where we follow the convention that summation is implied for repeated indices. By symmetry, $\delta T_{i}$ and $\delta T_{ijk}$ can be computed by $3$ and $10$ FFTs, respectively.

\subsection{Prescription of FFT}
\label{Sec:PrescriptionOfFFTs}

The FFT algorithm requires the interpolation of functions on a regular grid in position space. For any function $F(\vec{s}\,)$, the interpolation over the grid can be mathematically described in terms of the mass assignment function $W_{\rm mass}(\vec{s}\,)$ and the sampling function $\Pi(\vec{s}\,)$ as follows
\begin{eqnarray}
	\widetilde{F}_1(\vec{s}\,) = \Pi(\vec{s}\,) \int d^3s'\, W_{\rm mass}(\vec{s} - \vec{s}\,') F(\vec{s}\,').
	  \label{Eq:grid}
\end{eqnarray}
In the above expression, the sampling function is defined as
\begin{eqnarray}
	  \Pi(\vec{s}\,) = G^3 \sum_{\vec{n}} \delta_{\rm D}( \vec{s} - {\vec G} ),
\end{eqnarray}
where ${\vec G} = \{G_x n_x, G_y n_y, G_z n_z\}$ denotes grid points, $\vec{n}$ is an integer vector, $G_{i=x,y,z}$ are the grid spacing, $G^3\equiv G_xG_yG_z$, and the summation is over all three-dimensional integer vectors. The Fourier transform of $\widetilde{F}_1(\vec{s}\,)$ is given by
\begin{eqnarray}
	  \widetilde{F}_1(\vec{k}) &=& \int_{V} d^3s\, e^{-i\vec{k}\cdot\vec{s}} \widetilde{F}_1(\vec{s}\,) \nonumber \\
	  &=& \sum_{\vec{n}} W_{\rm mass}(\vec{k} - \vec{k}_{\rm f}) F(\vec{k} - \vec{k}_{\rm f}),
	  \label{Eq:grid_fourier}
\end{eqnarray}
where $\vec{k}_{\rm f} = (2\pi)\{n_x/G_x , n_y/G_y , n_z/G_z \}$, and $W_{\rm mass}(\vec{k})$ and $F(\vec{k})$ are the Fourier transforms of $W_{\rm mass}(\vec{s}\,)$ and $F(\vec{s}\,)$, respectively. Thus, the finite sampling results in the ``alias'' sums: i.e., the sums over $\vec{n}$~\citep{Jing2005ApJ...620..559J}.

To reduce the aliasing contribution, we adopt a simple technique proposed by~\citep{HockneyEastwood1981,Sefusatti2016MNRAS.460.3624S} that is based on the interlacing of two grids. The method significantly reduces the aliasing effect.  \citet{Sefusatti2016MNRAS.460.3624S} has not shown corrections for the shot-noise term, but the shot-noise term does not appear in the pairwise kSZ power multipoles (equation~\ref{Eq:estimator1}); therefore, we can apply this method to our kSZ analysis directly. We first prepare $\widetilde{F}_1(\vec{k})$ obtained by equation~(\ref{Eq:grid_fourier}). Second, we perform the interpolation on a grid shifted by the distance ${\vec G}/2$
\begin{eqnarray}
	  \widetilde{F}_2(\vec{k}) &=&  \int_{V} d^3s\, e^{-i\vec{k}\cdot\vec{s}}\, \Pi\left( \vec{s}+\frac{{\vec G}}{2} \right) \int d^3s'\,
	  W_{\rm mass}(\vec{s} - \vec{s}\,') F(\vec{s}\,') \nonumber \\
	  &=& \sum_{\vec{n}}\, (-1)^{n_{xyz}}\, W_{\rm mass}(\vec{k} - \vec{k}_{\rm f}) F(\vec{k} - \vec{k}_{\rm f}),
	  \label{Eq:F2}
\end{eqnarray}
where $n_{xyz}=n_x+n_y+n_z$. As a simple method to compute $\widetilde{F}_2(\vec{k})$, we compute $F(\vec{s}\,)$ for spatial galaxy positions shifted by the distance $\vec{G}/2$~\citep{SaitoInPrep}
\begin{eqnarray}
	  F'(\vec{s}\,) =  F\left( \vec{s}-\frac{\vec{G}}{2} \right)
	  = \sum_i^N F_i \delta_{\rm D}\left( \vec{s} - \vec{s}_i - \frac{\vec{G}}{2} \right),
	  \label{Eq:grid_dash1}
\end{eqnarray}
where $F(\vec{s}\,) = \sum_i^{N} F_i \delta_{\rm D}\left( \vec{s} - \vec{s}_i \right)$ with $F_i$ being a weight at galaxy $i$. Then, we can show that the Fourier transform of $F'(\vec{s}\,)$ in equation~(\ref{Eq:grid_fourier}) is related to $\widetilde{F}_2(\vec{k})$ through
\begin{eqnarray}
	  \widetilde{F}_2(\vec{k}) = e^{\frac{i}{2}\vec{k}\cdot\vec{G}} \widetilde{F}_1'(\vec{k}).
	  \label{Eq:grid_dash2}
\end{eqnarray}
Finally, we derive
\begin{eqnarray}
	  \widetilde{F}(\vec{k}) &=&  \frac{1}{2}\left( \widetilde{F}_1(\vec{k}) + \widetilde{F}_2(\vec{k}) \right) \nonumber\\
	  &=& \frac{1}{2}\sum_{\vec{n}} \left( 1 + (-1)^{n_{xyz}} \right) W_{\rm mass}(\vec{k} - \vec{k}_{\rm f}) F(\vec{k} - \vec{k}_{\rm f}).
	  \hspace{1cm}
	  \label{Eq:IR}
\end{eqnarray}
Thus, all the aliasing contributions correspond to odd values of the sum $n_x+n_y+n_z$, and can be partially corrected. In particular, the largest contributions for $|\vec{n}|=1$ are removed.

The Fourier transform of a function $F(\vec{s})$ measured by FFT, $F(\vec{k})|_{\rm FFT}$, incldues the effect of the mass assignment function $W_{\rm mass}(\vec{k}\,)$~\citep{Jing2005ApJ...620..559J}. We can remove such effects from $F(\vec{k})|_{\rm FFT}$ by simply dividing by $W_{\rm mass}(\vec{k})$: $F(\vec{k}) = F(\vec{k})|_{\rm FFT}/W_{\rm mass}(\vec{k})$. The most popular mass assignment function is given by
\begin{eqnarray}
	  W_{\rm mass}(\vec{k}) = \prod_{i=x,y,z}\left[ {\rm sinc}\left( \frac{\pi k_i}{2 k_{{\rm N},i}} \right) \right]^{p},
\end{eqnarray}
where $k_{{\rm N},i}=\pi/H_i$ is the Nyquist frequency of $i$-axis with the grid spacing $H_i$ on the axis. The indexes $p=1$, $p=2$, and $p=3$ correspond to the nearest grid point (NGP), cloud-in-cell (CIC), and triangular-shaped cloud (TSC) assignment functions, respectively.

\section{Derivation of equation~(53)}
\label{Ap:Derivations}

A Hankel transform relates the pairwise kSZ power spectrum estimator averaged in spherical shell of $\vec{k}$ (equations~\ref{Eq:estimator1} and \ref{Eq:avPk}) to the pairwise kSZ correlation function estimator normalized by the constant factor $A$ (equation~\ref{Eq:A}):
\begin{eqnarray}
	\widehat{P}_{\rm kSZ,\, \ell}(k)
	= 4\pi (-i)^{\ell} \int ds\, s^2 j_{\ell}(ks)\, \widehat{\xi}_{ {\rm kSZ},\, \ell}(s)
	\label{Ap:Hankel}
\end{eqnarray}
where 
\begin{eqnarray}
	\scalebox{0.85}{$\displaystyle
	\widehat{\xi}_{ {\rm kSZ},\, \ell}(s)
	$}
	&=& 
	\scalebox{0.85}{$\displaystyle
	 - \frac{\left( 2\ell+1 \right)}{A}
	\int \frac{d\Omega_s}{4\pi} \int d^3s_1 \int d^3s_2 \delta_{\rm D}\left( \vec{s} - \vec{s}_{12} \right) $} \nonumber \\
	&\times&
	\scalebox{0.85}{$\displaystyle
	\hspace{-0.3cm}{\cal L}_{\ell}(\hat{s}_{12}\cdot\hat{n}_{12}) 
	\Big[ \delta T(\vec{s}_1) \delta n(\vec{s}_2) - \delta n(\vec{s}_1)\delta T(\vec{s}_2) \Big].$}
	\label{Eq:estimator_xi}
\end{eqnarray}

A way to construct a theoretical model to explain what we observe, $\widehat{P}_{ {\rm kSZ},\, \ell}(k)$ (equation~\ref{Eq:estimator1}), is to compute its ensemble average, $\langle \widehat{P}_{ {\rm kSZ},\, \ell}(k)\rangle$, which may include survey geometry effects. The observed galaxy and kSZ fields, respectively $\delta n$ and $\delta T$, are multiplied by a mask that accounts for both the survey geometry and the completeness weight (equation~\ref{Weight}). The ensemble average of the square bracket in equation~(\ref{Eq:estimator_xi}) is given by 
\begin{eqnarray}
	&& \left\langle \delta T(\vec{s}_1) \delta n(\vec{s}_2) - \delta n(\vec{s}_1)\delta T(\vec{s}_2) \right\rangle \nonumber \\
	&=&  - \bar{n}( \vec{s}_1)\bar{n}( \vec{s}_2) \sum_{\ell}\xi_{ {\rm kSZ},\, \ell}(|\vec{s}_{12}|) {\cal L}_{\ell}(\hat{s}_{12}\cdot\hat{n}_{12}),
\end{eqnarray}
where the number density computed from a random catalogue $\bar{n}(\vec{s}_1)$ given by equation~(\ref{Eq:n_bar}) corresponds to the survey mask. Using the relation,
\begin{eqnarray}
	{\cal L}_{\ell}(\mu) {\cal L}_{\ell_1}(\mu)
	= \sum_{\ell_2} (2\ell_2+1)\left( \begin{smallmatrix} \ell & \ell_1 & \ell_2 \\ 0 & 0 & 0 \end{smallmatrix} \right)^2 {\cal L}_{\ell_2}(\mu),
\end{eqnarray}
we derive
\begin{eqnarray}
	\left\langle \widehat{\xi}_{ {\rm kSZ},\, \ell}(s)\right\rangle
	= (2\ell+1)\sum_{\ell_1\ell_2}  \left( \begin{smallmatrix} \ell & \ell_1 & \ell_2 \\ 0 & 0 & 0 \end{smallmatrix} \right)^2 
	\xi_{ {\rm kSZ},\, \ell_1}(s) Q_{\ell_2}(s),
	\label{Ap:xi_Q}
\end{eqnarray}
where 
$\left( \begin{smallmatrix} \ell_1 & \ell_2 & \ell \\ m_1 & m_2 & m \end{smallmatrix} \right)$ is the Wigner 3\textit{j} symbol, and the survey window function multipoles $Q_{\ell}(s)$ are given by 
\begin{eqnarray}
	Q_{\ell}(s) &=& \frac{\left( 2\ell+1 \right)}{A}
	\int \frac{d\Omega_s}{4\pi} \int d^3s_1 \int d^3s_2 \delta_{\rm D}\left( \vec{s} - \vec{s}_{12} \right) \nonumber \\
	&\times&{\cal L}_{\ell}(\hat{s}_{12}\cdot\hat{n}_{12})  \bar{n}(\vec{s}_1)\bar{n}(\vec{s}_2).
\end{eqnarray}
In the local plane parallel approximation, $\hat{n}_{12}\sim\hat{s}_1$, we have
\begin{eqnarray}
	Q_{\ell}(s) = \frac{\left( 2\ell+1 \right)}{A}
	\int \frac{d\Omega_s}{4\pi} \int \frac{d^3k}{(2\pi)^3} e^{i\vec{k}\cdot\vec{s}} \bar{n}_{\ell}(\vec{k}) \bar{n}^*(\vec{k}),
	\label{Eq:Q2}
\end{eqnarray}
where $\bar{n}(\vec{k})$ is the Fourier transform of $\bar{n}(\vec{s})$, and $\bar{n}_{\ell}(\vec{k})$ can be obtained by 
\begin{eqnarray}
	\bar{n}(\vec{k}) &=& \int d^3s e^{-i\vec{k}\cdot\vec{s}} \bar{n}(\vec{s}) \nonumber \\
	\bar{n}_{\ell}(\vec{k}) &=& \int d^3s e^{-i\vec{k}\cdot\vec{s}} {\cal L}_{\ell}(\hat{k}\cdot\hat{s})\bar{n}(\vec{s}).
\end{eqnarray}

Substituting equation~(\ref{Ap:xi_Q}) into equation~(\ref{Ap:Hankel}) results in equation~(\ref{Eq:xi_Q}). In this work, we do not take into account the integral constant correction~\citep{Peacock1991MNRAS.253..307P}, because we expect that the effect would be a sub-dominant contribution to the corrections from the survey window function. We leave a more detailed discussion of the effect as future works.

\bsp	
\label{lastpage}

\end{document}